\documentclass[sigconf]{acmart}
\usepackage{tabularx}
\usepackage{booktabs}
\setcopyright{none}
\AtBeginDocument{%
  }

\begin{document}

\title{Looks Fine to Me: Understanding Oversight in Agentic Interactions}

% \correspondingauthor
% \affiliation{%
\author{Dinithi Dissanayake}

\orcid{0009-0007-0178-984X}
\affiliation{%
  \institution{Augmented Human Lab \\ National University of Singapore}
  \city{Singapore}
  \country{Singapore}}
\email{dinithid@u.nus.edu}

\author{Yasith Samaradivakara}
\affiliation{%
  \institution{MIT Media Lab\\ MIT}
  % \city{National University of Singapore}
  \city{Cambridge, Massachusetts}
  \country{USA}}
\email{yasith@media.mit.edu}

\author{Hyung Woon Lee}
\affiliation{%
  \institution{MIT Media Lab\\ MIT}
  % \city{National University of Singapore}
  \city{Cambridge, Massachusetts}
  \country{USA}}
\email{hyung98@media.mit.edu}

\author{Yadeesha Weerasinghe}
\affiliation{%
  \institution{Augmented Human Lab \\ National University of Singapore}
  % \city{National University of Singapore}
  \city{Singapore}
  \country{Singapore}}
\email{yadeesha@ahlab.org}

\author{Sanjay Saha}
\affiliation{%
  \institution{Augmented Human Lab \\ National University of Singapore}
  % \city{National University of Singapore}
  \city{Singapore}
  \country{Singapore}}
\email{sanjaysaha@nus.edu.sg}

\author{Pattie Maes}
\affiliation{%
  \institution{MIT Media Lab\\ MIT}
  % \city{National University of Singapore}
  \city{Cambridge, Massachusetts}
  \country{USA}}
\email{pattie@media.mit.edu}

\author{Suranga Nanayakkara}
\affiliation{%
  \institution{Augmented Human Lab \\ National University of Singapore}
  % \city{National University of Singapore}
  \city{Singapore}
  \country{Singapore}}
\email{suranga@ahlab.org}
% \affiliation{%

% \affiliation{%

% \affiliation{%

% \affiliation{%

% \affiliation{%

% \affiliation{%

% \correspondingauthor
% \affiliation{%

\renewcommand{\shortauthors}{Trovato et al.}
\renewcommand\footnotetextcopyrightpermission[1]{}

\begin{abstract}
Agentic AI systems are deployed to execute complex tasks across domains such as writing, coding, and data analysis. As execution is offloaded to AI agents, the human role shifts toward oversight. Effective oversight requires recognizing when closer scrutiny is needed, even though flawed outputs may appear convincing. Our formative study (\(N=12\)) revealed that users accepted plausible outputs without thorough evaluation, and only recognized these lapses in retrospect. To examine this systematically, we conducted a controlled within-subject study (\(N=36\)) in which participants completed programming tasks with an agent that produced correct outputs, silent defects, or visible failures. Combining task outcomes, interaction logs, gaze, physiological signals, and self-reports, we examined perceived, behavioral, and effective oversight. Participants were often highly confident while missing silent defects: they performed generic checks, failed to generate diagnostic evidence, or even misinterpreted defect evidence.
We discuss how studying oversight across multiple dimensions can support future work on timely, task-calibrated re-engagement.

% they had produced. We %interpret these findings through the lens of \textit{mindlessness in agentic interaction} and 
\end{abstract}




\begin{teaserfigure}
 \includegraphics[width=\textwidth]{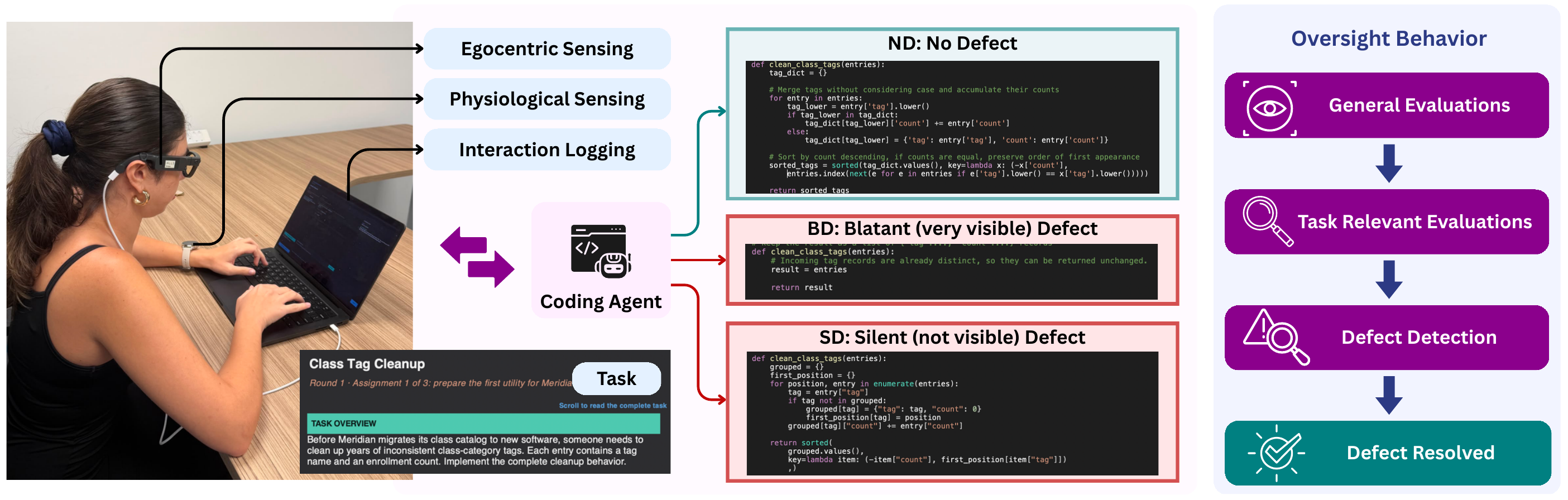}
 \centering
  \caption{Participants completed Python programming tasks with a coding agent while we collected egocentric sensing (FPV, gaze), physiological sensing (PPG, HRV), and interface-interaction logs (cursor movements, keyboard clicks, interface and agent interactions). Across tasks, the agent produced correct code (ND), a silent defect that passed the hidden pre-merge checks (SD), or a visible failure that caused the checks to fail (BD). We examined how participants progressed from surface-level checking to task-relevant evaluation, defect detection, and successful resolution.}
  \Description{A photograph on the left shows a participant wearing Project Aria glasses, typing on a laptop. Three labels point to the glasses and laptop: Egocentric Sensing, Physiological Sensing, and Interaction Logging. A double-headed arrow connects the participant to a Coding Agent icon and a task card titled ``Class Tag Cleanup.'' To the right, three bordered panels show excerpts of Python code for the three conditions: ND (No Defect), BD (Blatant, very visible Defect), and SD (Silent, not visible Defect). A fourth column, headed Oversight Behavior, lists four stages connected by downward arrows: General Evaluations, Task Relevant Evaluations, Defect Detection, and Defect Resolved.}
  \label{fig:teaser}
\end{teaserfigure}

\newcommand{\yasith}[1]{\textcolor{red}{#1}}
\newcommand{\hyung}[1]{\textcolor{orange}{#1}}

\received{20 February 2007}
\received[revised]{12 March 2009}
\received[accepted]{5 June 2009}

\maketitle
\renewcommand\footnotetextcopyrightpermission[1]{}
\pagestyle{plain}

% \newpage
\section{Introduction}
Generative AI is changing not only what people can do, but also how they are involved in the process. With a single prompt, AI agents can now interpret goals, plan and execute actions, modify artifacts directly, and return completed work. As users continue to offload task execution to AI agents, user roles are shifting towards oversight. However, ensuring that completed work aligns with user intent requires oversight that is active, not passive; users need to trace and understand the agent’s actions, and evaluate if the agent has performed their task correctly. 

Prior works discuss the necessity of remaining cognitively engaged, as past successes cannot guarantee future ones~\cite{chen2026code}. Repeated acceptance of outputs without thorough evaluation reduces opportunities for learning and critical thinking, while leaving users less prepared to catch subtle errors and failures, or intervene when an agent’s actions diverge from their intent ~\cite{lee2025impact,kazemitabaar2025exploring}. However, as agents become increasingly capable, checking every action becomes unnecessary and costly. Moreover, with repeated successes, it becomes harder for users to constantly remain cognitively engaged, and users may gradually fall out of the loop \cite{endsley1995out,parasuraman1993performance,bainbridge1983ironies,mccabe2024ironies}. We therefore employ the term \textit{effective oversight} to mean recognizing when closer evaluation is needed and re-engaging \textit{at the right moment}. Conversely, a \textit{lapse} in effective oversight would describe the phenomenon where users remain involved in an agentic workflow and accept outputs
%\textit{mindlessness in agentic interaction}
% to describe the phenomenon where users remain involved in an agentic workflow and accept outputs %without evaluating details that should require human judgment. %We therefore view mindlessness as a breakdown of effective oversight. 
%Users may remain procedurally involved 
while insufficiently evaluating the substance of the agent's work. 

% We observed this behavior in our formative study (\(N=12\)) of everyday AI use, where participants go mindless even without realizing they are mindless. This creates a central challenge: \emph{How does oversight unfold during agentic interactions?} Our formative study identified coding as a potential scenario in which participants most readily accepted AI output with limited checking.  

Effective oversight is often a challenge for users, given that flawed outputs are often generated as complete and plausible responses.
Langer describes a similar failure to spot %flaws as \textit{mindlessness}: following familiar 
cues while overlooking differences that matter in the current situation \cite{langer1978mindlessness,langer1989minding}. 
This characterization of effective oversight raises questions on %beyond reliance outcomes alone: 
whether users recognize lapses in their oversight, how these lapses manifest during interaction, and what prompts them to re-engage. 
To investigate oversight in agentic interaction, we first conducted a formative study (\(N=12\)) on everyday AI use and discovered that lapses in effective oversight within agentic interactions often occur without the user even being aware of it. Guided by our findings, we identified AI-assisted programming as a prominent context to study oversight, and subsequently
conducted a controlled within-subject study with 36 participants completing programming tasks with an AI coding agent. The agent either produced a correct solution, introduced a plausible defect without an external failure signal, or introduced a defect surfaced by a visible failure. %Our findings show that participants often missed silent defects while remain confident, yet were able to easily identify faulty outputs when the failure was made visible, demonstrating that regardless of capability, effective oversight depended heavily on visible failure cues.
Our findings show that participants often missed silent defects despite remaining confident, but readily identified faulty outputs when failures were visible, suggesting that visible failure cues played an important role in prompting effective oversight.
%
% By combining task outcomes, interaction telemetry, repeated self-report, gaze, and physiological sensing, we characterize moment-to-moment patterns of oversight and examine how they vary across agent behaviour, repeated interactions, and users with different levels of programming experience.
% insights on how people  experienced reduced cognitive engagement with AI.
Overall, this paper makes two contributions:
\begin{itemize}
  \item \textbf{%A case for studying oversight where it happens, in the moment.
  Insights on how people experienced reduced oversight with AI.
  } Our formative study (\(N=12\)) shows that users can describe reduced oversight but may not recognize it as it happens. We share several themes spanning across interaction behaviour, attention, and bodily state, as well as explore and chart the directions this exploration opens.
    %
    %outline several research directions.
    %
   % The signs reported were diverse, spanning interaction behaviour, attention, and bodily state. We therefore outline several o explore how lapses in oversight appear in users’ moment-to-moment interactions and chart the directions this exploration opens.
  \item \textbf{Findings on `oversight' under repeated agent interactions based on multimodal data analysis.} 
    %
    %A multimodal portrait of oversight under repeated agent interaction.} 
  Through a controlled study (\(N=36\)) combining task outcomes, interaction logs, gaze, and physiological signals, we examine how oversight changes between interactions with correct, silently flawed, and visibly failing agent outputs, and how these patterns vary with user programming experience. We reveal differences between feeling vigilant and being vigilant and discuss their implications for measuring oversight and supporting timely re-engagement.
    % \item \textbf{Evidence that feeling vigilant and being vigilant come apart.} We identify a dissociation between confidence, perceived thoroughness, observable verification, and defect detection. We discuss what these differences mean for measuring oversight and designing agentic systems that support timely re-engagement. 
\end{itemize}

\section{Related Work}
\subsection{Cognitive Engagement and Oversight in Human--AI Interaction}
Concerns about maintaining human cognitive engagement have long accompanied increasing levels of automation. Bainbridge's ``ironies of automation'' highlighted how automating routine execution can shift people toward supervisory roles while reducing the experience needed to intervene effectively when automation fails~\cite{bainbridge1983ironies}. Subsequent work showed that reliable automation can foster complacency and reduced monitoring, while higher levels of automation can leave users ``out of the loop'' and less able to understand system state or intervene effectively~\cite{parasuraman1993performance,endsley1995out,parasuraman1997humans}. Related work on automation bias and reliance further shows that people may defer to automated recommendations rather than independently seek or evaluate information, with reliance shaped by factors such as trust, workload, and system reliability~\cite{skitka2000automation,dzindolet2003trust,lee2004trust,parasuraman2010complacency}. As systems take over more of the execution, the human role increasingly shifts toward oversight. Yet sustained attention and evaluation can become more difficult as continuous involvement in the task decreases.

Generative AI extends this shift from supervising automated processes to evaluating AI-generated work. As users delegate more cognitive work to AI, studies have raised concerns about cognitive offloading and reduced critical-thinking effort~\cite{gerlich2025ai,lee2025impact}. Delegation does not necessarily remove the need for human evaluation. Users must still decide what to delegate, assess generated outputs, and determine when further scrutiny is needed~\cite{tankelevitch2024metacognitive}. When such evaluation is insufficient, users may over-rely on AI and accept incorrect recommendations~\cite{buccinca2021trust, deJong2025cognitive}. This has motivated interventions such as cognitive forcing functions, which encourage users to engage more deliberately with a problem before or while receiving AI assistance~\cite{buccinca2021trust, deJong2025cognitive}.

However, much of this work characterizes cognitive engagement through its outcomes, such as reliance on incorrect advice or task performance~\cite{buccinca2021trust, deJong2025cognitive}, through users' retrospective reports of critical-thinking effort~\cite{lee2025impact}, or by evaluating interventions designed to increase deliberation~\cite{buccinca2021trust, deJong2025cognitive}. Less is understood about how the quality of oversight changes \emph{during ongoing interaction}, particularly when users remain behaviorally involved with an AI system and nothing explicitly signals that greater scrutiny is necessary. Human involvement alone does not necessarily imply effective oversight~\cite{zhu2026meaningful}: a user may continue prompting, reviewing, or approving an agent's actions while attending only superficially to details that require human judgment. We examine this in repeated interactions with increasingly autonomous AI agents.

%We draw on Langer's concept of \emph{mindlessness} to characterize this distinction. Mindlessness describes responding to familiar cues or established categories while overlooking distinctions that matter in the current context~\cite{langer1978mindlessness,langer1989minding}. In human--AI interaction, this suggests that users may remain procedurally involved while insufficiently evaluating the substance of an agent's work. This raises questions beyond reliance outcomes alone: whether users recognize lapses in their oversight, how these lapses manifest during interaction, and what prompts them to re-engage. We examine these questions in repeated interactions with increasingly autonomous AI agents.

% We draw on Langer's concept of \emph{mindlessness} to characterize this distinction. Mindlessness describes responding to familiar cues or established categories while overlooking distinctions that matter in the current context~\cite{langer1978mindlessness,langer1989minding}. In human--AI interaction, this suggests that users may remain procedurally involved while insufficiently evaluating the substance of an agent's work. This raises questions beyond reliance outcomes alone: whether users recognize lapses in their oversight, how these lapses manifest during interaction, and what prompts them to re-engage. We examine these questions in repeated interactions with increasingly autonomous AI agents.

\subsection{From AI-Assisted Coding to Agentic Programming}
AI-assisted programming has evolved from supporting developers during coding to increasingly delegating programming tasks to AI systems. Code-completion tools operate within developer-led workflows, leaving programmers to inspect, accept, modify, or reject generated suggestions~\cite{mozannar2024reading}. More recent coding agents can instead take higher-level goals, plan changes, modify files, execute code and tests, and iterate toward solutions with substantially less developer involvement~\cite{takerngsaksiri2025human,chen2026code}. This progression represents not simply an increase in generation capability, but a redistribution of work between developers and AI systems.

%As execution is increasingly delegated, the developer's role shifts toward supervising and verifying autonomously produced work. 
Studies of agentic development workflows show that agents can take substantial responsibility for execution while humans retain responsibility for reviewing, steering, and accepting their work~\cite{chung2026collaborator,wang2026humans,dhanorkar2026oversight}. Yet verification can itself be demanding. Developers must evaluate implementations they may not have directly produced or closely followed~\cite{mccabe2024ironies}, and repeated verification can impose cognitive load and contribute to fatigue~\cite{fan2026help}. Prior work has therefore examined how developers critically engage with generated code and how deeper engagement might be supported~\cite{hutchison2026tab,kazemitabaar2025exploring}. At the same time, studies of AI-assisted programming raise concerns that increased generation may not be accompanied by equally rigorous validation~\cite{obrien2025more}. As coding agents become more autonomous, effective oversight increasingly depends on how developers evaluate work they did not directly produce.

This challenge is particularly consequential when an agent appears to have completed a task successfully. Real-world coding-agent interactions reveal failures arising from mismatches between what developers intend and what agents interpret, implement, or report~\cite{tang2026coding}. Moreover, developers can have difficulty distinguishing correct from incorrect AI-generated artifacts even while remaining confident in their judgments~\cite{kaufman2026programmers}. Some failures provide clear cues for further investigation, such as failed tests, execution errors, or visibly incorrect results. When an implementation appears plausible and available checks pass, however, users must independently determine whether the evidence of success is sufficient. Less is understood about whether developers recognize when such apparent success warrants deeper scrutiny and whether their evaluation can reveal errors that are not externally signaled.

\subsection{Modeling Cognitive Engagement from Behavior}
% Cognitive states are not directly observable, but can manifest through patterns of behavior during interaction. While cognitive states are not directly observable prior work has therefore modeled states such as attention, engagement, confusion, and cognitive load from observable signals produced while people perform tasks~\cite{li2021measuring,TODO}. These signals span interaction behavior, visual attention, and physiological responses. Eye-tracking measures such as fixations, saccades, dwell time, gaze transitions, and pupil responses have been used to characterize attention and cognitive processing~\cite{TODO_eye_review}, while physiological measures including heart rate, heart-rate variability, electrodermal activity, and respiration have been used to characterize cognitive workload, arousal, and mental effort~\cite{tierney2025cognitive,TODO_physio_review}. Behavioral measures such as pauses, response timing, typing activity, mouse movement, and action sequences provide another view into how cognitive processes unfold through interaction~\cite{TODO_behavior}. Together, these modalities provide complementary, though indirect, evidence about users' cognitive engagement during task performance.
Understanding the quality of human oversight requires looking beyond whether users remain active in an interaction. Prior work suggests that differences in cognitive engagement can be reflected in how people behave, allocate attention, and respond physiologically while performing a task~\cite{li2021measuring,dmello2010multimodal,kosch2023workload}. Behavioral measures such as pauses, response timing, and action sequences can reflect changes in how users approach a task~\cite{dmello2009body,stiber2024uh}; gaze measures provide evidence about the allocation of visual attention; and physiological measures have been studied as indicators of workload, arousal, and mental effort~\cite{tao2019physiological,charles2019mental,kosch2023workload}. These signals offer potential evidence about cognitive engagement, but their meaning is rarely unambiguous. The same behavior or physiological response may reflect different cognitive processes depending on the task and moment in which it occurs. Cognitive engagement is therefore better characterized by considering multiple signals and the context in which they occur, rather than interpreting individual measures in isolation~\cite{li2021measuring}.

This need for context is particularly important in Human--AI interaction, where users' actions can provide task-grounded evidence of how they engage with AI-generated outputs. Interaction traces such as prompting, accepting or rejecting suggestions, editing, testing, and sequences of actions have been used to characterize how developers interact with and evaluate generated code~\cite{mozannar2024reading,hutchison2026tab}. More broadly, sequences of programming actions can reveal strategies that are difficult to infer from individual low-level behaviors alone~\cite{fern2007mining}. For oversight, however, observable activity does not necessarily imply substantive evaluation. Continued interaction may reflect careful inspection, but it may also consist of routine actions that provide little evidence that the agent's work has been critically examined. Conversely, periods of limited interaction may involve careful reading or reasoning. Thus, interpreting behavioral traces as evidence of oversight requires considering not only \emph{whether} users interact, but \emph{what} they are doing and how those actions relate to the task and the agent's work.

Self-report provides a complementary perspective by capturing users' perceived effort, workload, confidence, and engagement~\cite{greene2015selfreport,li2021measuring}. Yet subjective, behavioral, physiological, and performance-based measures need not converge~\cite{charles2019mental,kosch2023workload}. This distinction is particularly important for human oversight of AI agents: users may perceive themselves as attentive and exhibit behaviors consistent with evaluation without actually scrutinizing the aspects of an agent's work that require human judgment. Effective oversight therefore cannot be inferred from perceived or observable engagement alone. Understanding oversight requires considering these indicators alongside whether users perform task-relevant evaluation capable of revealing problems in the agent's work. This motivates examining oversight across perceived, behavioral, and effective dimensions, and asking when these dimensions align---and when they diverge.

\section{Formative Study}
We conducted a formative interview study to understand how people experience changes in their oversight during everyday AI use. We aimed to identify when users reduce oversight, whether they recognize this change, and how it might manifest in observable behavior. This formative study provided an empirical basis for selecting the context, manipulations, and measurements used in our controlled study.

\subsection{Participants \& Procedure}

We conducted semi-structured interviews with 12 regular users of generative AI $((M_{\mathrm{age}}=27.7)$; 5 Females and 7 Males). Participants included four undergraduate students (F2, F3, F7, F10), four graduate students (F5, F6, F11, F12), and four working professionals ([F1, F4, F8, F9]). Participants used generative AI for activities including programming, writing, learning, and information seeking. Participants first described how they used AI in their everyday work. We then asked them to recall situations in which they had accepted or acted on AI output without carefully evaluating it. Follow-up questions examined what contributed to these situations, whether participants recognized them as they occurred, and what changes they noticed in their interaction, attention, or bodily state. Participants also reflected on the benefits and risks of reduced engagement and whether an external cue might help them re-engage. The full interview protocol is provided in the Appendix Section~\ref{sec:app:formativeprotocol}.

% We first coded excerpts describing participants' AI-use practices, experiences of reduced evaluation, its perceived triggers and consequences, observable signs, and awareness of these changes. We iteratively compared related codes across participants and organized them into higher-level themes. The resulting themes informed the design of our subsequent controlled study.
We used a primarily deductive coding approach, with broad analytic categories derived from the interview guide. We coded excerpts describing participants' AI-use practices, experiences of reduced evaluation, perceived triggers and consequences, observable signs, and awareness of these changes. The research team regularly discussed interviews from the initial participants to refine the interview guide, identify recurring codes, and clarify code definitions. The first author then coded the full dataset, while a second author reviewed the resulting coding. We compared related codes across participants and organized them into higher-level patterns. Below, we describe these patterns and how they informed the design of our subsequent controlled study.

\subsection{Findings}

\subsubsection*{\textbf{Oversight was calibrated to its perceived value}}
Participants did not consider close evaluation necessary in every interaction. Instead, they adjusted their oversight according to the perceived stakes, available time, interest in the task, and confidence in the AI. Under time pressure, participants prioritized completing the task over understanding how the output had been produced. F5 explained, \emph{``When I am rushing, I am more mindless because I just need to get it done, even if it is lower quality."} Similarly, low-stakes tasks gave participants little reason to invest in verification. As F8 described, \emph{``If the task is high-priority or high-value \ldots I will not directly use the output the AI gives me,"} whereas low-stakes outputs were more readily accepted.

Repeated success also reduced the perceived value of continued checking. F1 described initially examining the system's work but later deciding, \emph{``Once I know it is doing a good job, then I become mindless. I think, Okay, I can offload this. Let it do it."} Thus, reduced oversight was not always an indiscriminate tendency to trust AI. It was often a contextual response to the apparent costs and benefits of further evaluation.

Participants also recognized benefits to this offloading. It saved time, reduced effort, and allowed them to focus their attention elsewhere. However, they described a longer-term tradeoff involving learning and skill. F11 explained, \emph{``If we do not use AI tools, we will be behind, because others are using AI tools \ldots It is a tradeoff."} F12 similarly felt that AI made work faster but that \emph{``my coding skills are getting rusty because I generate everything from AI."} These accounts distinguish efficient delegation from the loss of evaluation needed to learn from, take responsibility for, or intervene in the work.

\subsubsection*{\textbf{Procedural interaction did not guarantee evaluation}}

Participants described remaining active in an AI interaction without necessarily reasoning about its content. They could continue prompting, approve actions, or repeatedly ask the AI to fix a problem while focusing only on whether the final result worked. F3 explained, \emph{``You know what the end goal is, and you know what the final result should be, but you do not care what commands or outputs the AI is giving as long as the end goal is achieved."} The same participant emphasized that even a long interaction could remain mindless when follow-ups consisted mainly of responses such as \emph{"yes," "no," or "this did not work, fix this."}

Agentic tools appeared to strengthen this separation between participation and evaluation. F3 contrasted earlier copy-and-paste interactions with coding agents that iterate independently: \emph{``Now, with Codex and similar tools, you ask it one shot and it does it. You do not have to go back and forth because they iterate on their own \ldots you do not think."} Participants commonly evaluated only the visible result. F2 described asking an agent to implement something, checking a few functions, and continuing without reading the generated code. F11 similarly stated, \emph{``I do not go through much of what it is doing unless it is not working."}

These accounts suggest that the amount of interaction is not, by itself, evidence of effective oversight. Users may remain procedurally involved while attending only to surface indicators such as whether the agent has finished, whether the output appears plausible, or whether the application runs.

\subsubsection*{\textbf{Awareness ranged from deliberate offloading to retrospective recognition}}
Participants differed in whether they recognized their reduced engagement. Some intentionally chose not to evaluate low-priority work. F3 stated, \emph{``I know I am doing it. I know it is not good \ldots I do not think it is an awareness problem. It is more of a discipline problem."} F5 similarly described remaining in control and intentionally becoming less engaged when the task did not warrant greater effort.

Other accounts indicated that reduced oversight could occur without being noticed at the time. F12 explained, \emph{"If I am really busy, I might not even realize that I am doing it mindlessly."} Several participants described recognizing the problem only when an output violated their expectations. F6 recalled that after an AI cited an article that did not exist, the failure \emph{"snapped me out of it"} and served as a reminder \emph{"to put in more effort, verify things, and check things."}

These findings complicate any simple claim that users are either aware or unaware of lapses in effective oversight. Awareness was contextual: users could deliberately offload evaluation in one situation but fail to notice inadequate oversight in another. Importantly, a visible failure could expose this mismatch and reactivate scrutiny. Whether that re-engagement persists beyond the immediate failure, however, remained unclear.

\subsubsection*{\textbf{Reduced oversight had diverse observable signs}}
Participants associated reduced oversight with changes in how they prompted, read, verified, and distributed their attention. Short or underspecified prompts were mentioned frequently. F1 explained, \emph{``If I am thinking about it, my prompt is long \ldots If I am mindless, I am more like, `Can you do this?'"} Participants also described fewer substantive follow-ups, limited external verification, skimming summaries, and approving commands without examining them.

Attention often shifted away while the agent worked. F2 described giving an agent a prompt, doing something else, and returning only to inspect what it had built. F5 similarly reported scrolling through YouTube while Codex worked in the background. Reported signs also extended beyond interaction logs: F3 associated reduced engagement with tiredness and slouching.

No single behavior consistently represented effective or ineffective oversight. For example, short prompts could reflect either shallow engagement or an efficiently specified task, while extensive back-and-forth interaction could still consist of automatic correction requests. The diversity of reported signs therefore motivated examining oversight through multiple signals rather than presupposing a single behavioral marker.

\subsection{Implications for the Controlled Study}

The formative study was conducted with a limited pool of participants, and as such, a further controlled study was necessary to better examine how effective oversight might change during and across agentic interactions. Nevertheless, the formative study helped guide the design of our controlled study in four ways. First, participants most readily described reduced evaluation in AI-assisted programming. Coding agents could autonomously produce complete, executable outputs, while the correctness of those outputs could be objectively evaluated in the case of testable code generation. We therefore selected programming as a controlled setting in which behavioral oversight and task outcomes could be examined together.

Second, participants often relied on visible functionality to decide whether further evaluation was needed. We therefore designed two failure types that differed in whether the incorrect solution produced an external failure signal. Silent failures let us examine whether users initiated scrutiny independently, whereas visible failures let us examine how oversight changed when the need for human judgment was made explicit.

Third, participants described failures as events that could \emph{snap} them back into active evaluation. We therefore studied repeated interactions to determine whether re-engagement remained limited to the visible failure or carried into subsequent tasks.

Finally, the reported signs of reduced oversight spanned interaction behavior, visual attention, bodily state, and subjective experience. We therefore captured multiple complementary measures in our main study: task outcomes to assess oversight effectiveness, interaction telemetry and gaze to capture observable behavior and attention, physiological signals to reflect bodily responses, and repeated self-reports to assess participants’ perceived engagement.

\section{AI-assisted Programming Study}
%In this section, we describe the main study design, participants, interface, and procedure.
% to examine how people oversee agentic interactions. Participants completed six Python programming tasks framed as implementing six features for a company. They worked at their own pace with the assistance of an autonomous coding agent.
Guided by our formative findings, we designed a controlled laboratory study in which, we examined how users’ oversight changes across repeated interactions with a coding agent and whether a visible failure prompts them to re-engage more closely. We combined behavioral, gaze, physiological, and self-report measures to capture different aspects of their engagement. Smart glasses (Meta Aria Gen 2 glasses\footnote{\url{https://www.projectaria.com/}}) were used to record gaze and egocentric activity during these interactions, while a Garmin watch was used to record physiological data. Figure~\ref{fig:teaser} explains the basic study setup.

% The formative interviews raised an intriguing possibility: interacting with coding agents may reduce puser oversight behavior-mainly to stud oversightbehavior. Curious to examine how this unfolds during actual use, we designed an exploratory controlled study in which participants repeatedly interacted with a coding agent. We combined behavioral, gaze, physiological, and self-report measures to capture different aspects of their engagement. Project Aria glasses were used only to record gaze and egocentric activity during these interactions.

\subsection{Participants}
We recruited 44 participants who met two inclusion criteria: (1) self-reported basic proficiency in Python and (2) the ability to work comfortably on a laptop while wearing the Aria glasses without prescription glasses. 
% Participants who normally wore prescription glasses could participate if they were comfortable working without them or wore contact lenses during the study; several participants chose the latter option.
Six were run as an initial pilot block and excluded from analysis as we changed the UX/UI interactions and task setup based on their feedback; two additional participants were excluded due to incomplete sessions (application/recording failures partway through the pipeline), leaving a final analytic sample of \textbf{36 participants}. Participants ranged in age from 19 to 35 ($M = 25.4$) and reported their gender as male (23), female (12), or non-binary (1). Self-rated Python proficiency was predominantly intermediate (23 participants comfortable with functions, data structures, and common libraries), with 6 beginners (able to read and write simple scripts) and 7 advanced programmers (able to write complex programs comfortably). The study was approved by the Departmental Ethics Review Committee at [XXX University]. All participants provided informed consent and were compensated for their time.

\subsection{Programming Interface}

\begin{figure*}[h]
    \centering
    \includegraphics[width=\textwidth]{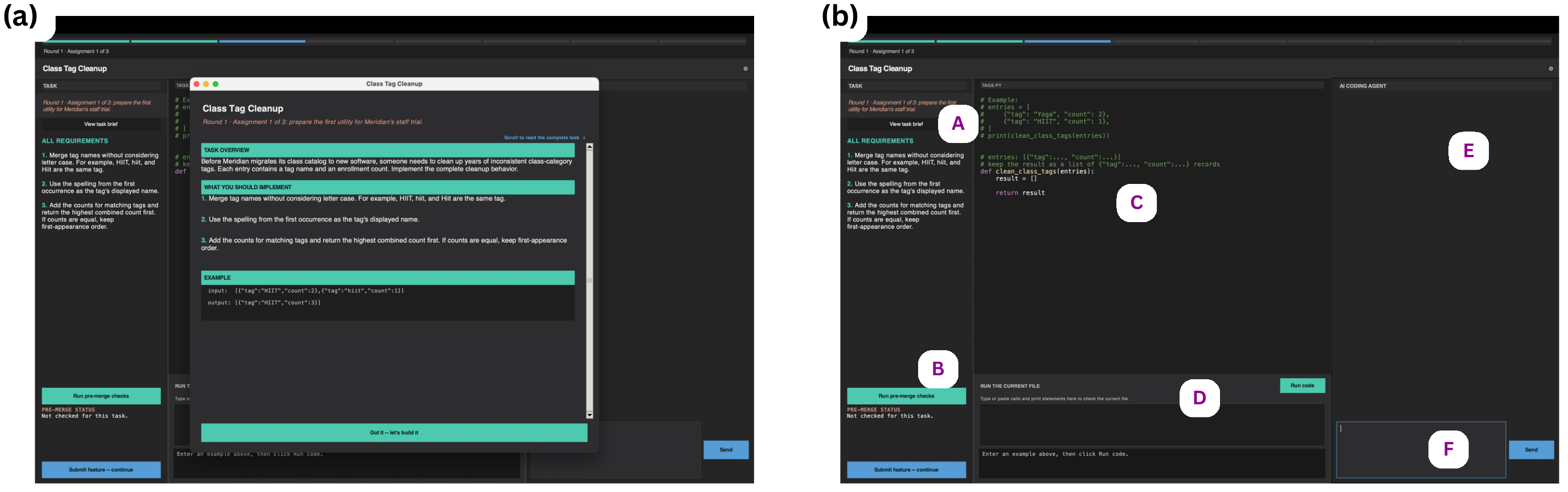}
    \caption{The programming interface used in the main study. \textbf{(a):}
Participants could open the full task brief to review the task overview,
requirements, and example input and output. Selecting ``Got it---let's build
it!'' returned them to the main workspace. \textbf{(b):} The main workspace
provided (A) access to the task brief, (B) the provided pre-merge checks,
(C) the code editor, (D) a panel for manually executing the code with custom
inputs, (E) the coding agent's responses, and (F) an input field for prompting
the agent and asking follow-up questions.}
\Description{Two screenshots of the study interface. On the left, the full task brief opens as a panel over the workspace, listing the task overview, requirements, and an example input and output, with a ``Got it---let's build it!'' button at the bottom. On the right, the main workspace is shown with six regions labeled A to F: (A) the task brief button, (B) the provided pre-merge checks, (C) the code editor containing scaffolded Python, (D) a panel for running the code with custom inputs, (E) the coding agent's responses, and (F) a text field for prompting the agent.}

    \label{fig:thecodinginterface}
\end{figure*}

We developed a desktop study interface that combined a task brief, a Python code editor, and an AI coding agent (Refer to Figure~\ref{fig:thecodinginterface}). The task brief presented all requirements for the current feature and could be reopened at any time. The editor contained scaffolded code with syntax highlighting and supported normal indentation and direct editing. Participants could also execute their own examples in a separate \textit{Run code} area. The coding agent operated autonomously: when participants requested a change, the agent's edit was applied directly to the file. Participants could subsequently inspect, modify, test, or discuss the resulting code with the agent.

Each task included hidden pre-merge checks. Participants were required to use the coding agent and obtain a passing result before handing off the feature. However, the introductory instructions explained that the checks covered only basic functionality and did not guarantee that every requirement or edge case was correct. This distinction allowed us to examine whether participants treated a passing check as sufficient evidence or performed additional verification through code inspection, customized examples, or follow-up questions. %The interface logged prompts, agent responses, code changes, manual executions, pre-merge results, task-brief use, focus changes, and handoff timing.

\subsection{Procedure}
\label{sec:study-design}
% Guided by our formative findings, we designed a controlled laboratory study% to examine how people oversee agentic interactions. Participants completed six Python programming tasks framed as implementing six features for a company. They worked at their own pace with the assistance of an autonomous coding agent.
% in which we examined how users’ oversight changes across repeated interactions with a coding agent and whether a visible failure prompts them to re-engage more closely.

%For each task, participants used the agent to implement a specified function within scaffolded Python code. The agent autonomously generated and immediately applied code edits. Participants could prompt the agent, ask follow-up questions, inspect or manually edit the code, run their own examples, and execute the provided pre-merge checks. They were required to use the agent and pass these checks before handing off each feature.

\begin{table}[t]
\centering
\caption{Experimental conditions. Outcomes describe the initial
agent-generated implementation before participant intervention.}
\label{tab:defect-conditions}
\small
\begin{tabularx}{\columnwidth}{@{}l c c X@{}}
\toprule
\textbf{Condition} &
\textbf{Code} &
\textbf{Checks} &
\textbf{Defect visibility} \\
\midrule
No Defect (ND) &
Correct & Pass &
No defect present. \\
\addlinespace

Silent Defect (SD) &
Incorrect & Pass &
Requires code inspection, reasoning, or targeted tests. \\
\addlinespace

Blatant Defect (BD) &
Incorrect & Fail &
Revealed by a comment accompanying failed checks. \\
\bottomrule
\end{tabularx}
\end{table}

% \begin{table}[t]
% \caption{Experimental conditions. Outcomes describe the initial agent-generated implementation before any participant intervention.}
% \label{tab:defect-conditions}
% \small
% \begin{tabularx}{0.5\columnwidth}{@{}l l l X@{}}
% \toprule
% \textbf{Condition} &
% \textbf{Implementation} &
% \textbf{Pre-merge checks} &
% \textbf{How the defect becomes visible} \\
% \midrule
% No Defect (ND) &
% Correct &
% Pass &
% No defect is present. \\
% \addlinespace

% Silent Defect (SD) &
% Incorrect &
% Pass &
% Only through code inspection, reasoning about the requirements, or additional targeted testing. \\
% \addlinespace

% Blatant Defect (BD) &
% Incorrect &
% Fail &
% Exposed by a comment accompanying the failed pre-merge checks. \\
% \bottomrule
% \end{tabularx}
% \end{table}

\subsubsection{Conditions} Participants used the agent to implement a specified function within scaffolded Python code. We manipulated whether the agent-generated implementation contained a defect and whether the visible pre-merge checks exposed it. Table~\ref{tab:defect-conditions} summarizes the three conditions. Both the Silent Defect (SD) and Blatant Defect (BD) implementations were incorrect; however, only the BD implementation caused the provided pre-merge checks to fail.

In the \textbf{No Defect (ND)} condition, the agent-generated implementation correctly satisfied the task requirements and passed the visible pre-merge checks. In the \textbf{Silent Defect (SD)} condition, the implementation violated a less apparent requirement or failed on an edge case while still passing the visible checks. Participants could therefore detect the defect only by carefully inspecting the code, reasoning about the requirements, or conducting additional targeted tests. In the \textbf{Blatant Defect (BD)} condition, the implementation contained an error that caused the visible pre-merge checks to fail. A comment accompanying the failed checks explicitly exposed the issue, providing participants with an external signal that the implementation required further attention.

\subsubsection{Tasks}
Participants first completed two short practice problems to learn the interface. The first introduced the coding agent, editor, manual execution area, and pre-merge checks through a simple arithmetic function. The second provided Python scaffolding for an easy-to-solve list-filtering problem and allowed participants to practice editing code directly without the help of an agent. This task allowed us to confirm that all participants had the basic Python knowledge needed to complete the study.

The main study consisted of six Python programming tasks, organized into two sets of three. In each task, participants used the coding agent to implement a single function. The task descriptions and corresponding silent defects are provided in Appendix~\ref{sec:app:protasks}. Participants could inspect and edit the generated code, execute it with their own inputs, ask the agent follow-up questions, and run the provided pre-merge checks. Participants proceeded to the next task after the pre-merge checks passed and they were satisfied with their solution.

Participants completed one task set in each of two rounds. Task-set order and the order of the no-defect, silent-defect, and blatant-defect conditions were counterbalanced across participants. The same condition order was repeated in the second round using the other task set. This design allowed us to compare oversight across defect conditions while reducing systematic effects associated with individual tasks and task-set order. After the first round, participants were shown the results for their first three solutions before continuing to the second round.

\begin{figure*}[h]
    \centering
    \includegraphics[width=\textwidth]{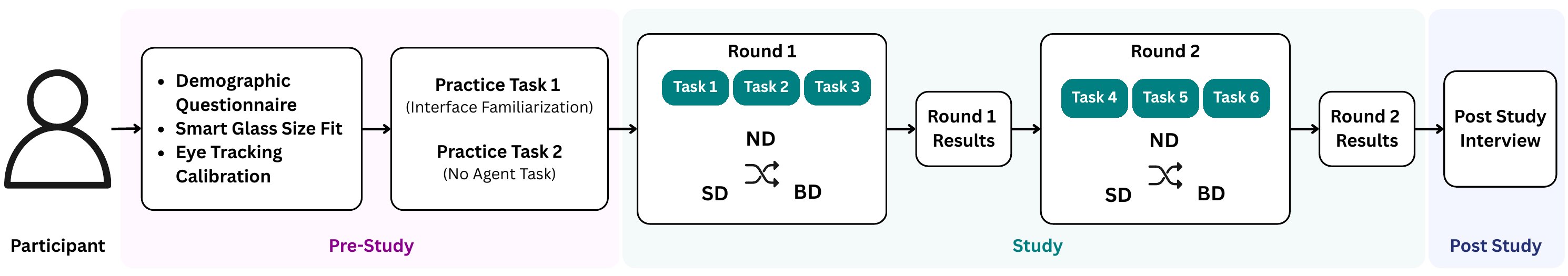}
    \caption{Main-study procedure. After demographics, smart glass fit and eye-tracking calibration, and two practice tasks, participants completed six programming tasks across two rounds. Each round contained one no-defect (ND), silent-defect (SD), and blatant-defect (BD) condition in randomized order. Participants received performance feedback after each round and completed a post-study interview.}
    \Description{A left-to-right flow diagram of the study session, divided into three phases. Pre-Study: a participant icon, then a box listing the demographic questionnaire, smart glass size fit, and eye-tracking calibration, then Practice Task 1 (interface familiarization) and Practice Task 2 (no agent task). Study: Round 1 contains Task 1, Task 2, and Task 3, with the ND, SD, and BD conditions shown below and a shuffle symbol indicating randomized order; Round 1 Results follows. Round 2 repeats the same structure with Task 4, Task 5, and Task 6, followed by Round 2 Results. Post Study: a post-study interview.}

    \label{fig:studydiagram}
\end{figure*}

%After each task, participants completed the NASA-TLX questionnaire and rated how thoroughly they had evaluated the agent's suggestions and how confident they were that their final solution met all requirements. After completing both rounds, participants took part in a five- to ten-minute semi-structured interview about their evaluation strategies, responses to the release outcomes, and any changes in their approach during the second round.
%
%\subsection{Procedure}
%
\subsection{Data Collection} We recorded participants’ prompts, agent responses, code changes, manual program executions, use of the pre-merge checks, access to the task brief, and task completion times. Participants wore Aria Gen~2 smart glasses to capture first-person video, gaze, and photoplethysmography (PPG). They also wore a Garmin smartwatch, which provided additional physiological measures. We separately recorded participants’ screens, keystrokes, and cursor movements. All data streams were synchronized with the study interaction logs. After each task, participants completed the NASA-TLX workload questionnaire. They also rated how thoroughly they had evaluated the agent’s suggestions and how confident they were that their final solution met all specified requirements.

At the end of the study, we conducted a brief semi-structured interview where %lasting five to ten minutes. 
participants reflected on how they interacted with the coding agent, evaluated its outputs, and results %responded to the release results 
presented between rounds. We asked whether these results affected their confidence or verification behavior, whether they changed their approach during the second round, and what consequences might arise from relying on the agent without sufficiently examining its work.

\section{Results}
% \section{Analysis and Results}
% \label{sec}
\begin{table}[t]
\centering
\caption{Definitions and operationalizations of the three dimensions of
oversight used throughout this paper.}
\label{tab:oversight_dimensions}
\begin{tabularx}{\columnwidth}{@{}p{0.2\columnwidth}XX@{}}
\toprule
\textbf{Dimension} & \textbf{Working definition} & \textbf{How we measure} \\
\midrule

\textbf{Perceived oversight}
& Participants' assessment of how carefully and successfully they evaluated
the agent's work.
& Self-reported confidence (1-5) and perceived thoroughness (1-5). \\

\textbf{Behavioral oversight}
& Observable actions through which participants sought or evaluated evidence
beyond the agent's initial output.
& Code inspection, requirement revisiting, manual execution, additional tests,
questions to the agent, and evaluation--revision loops. \\

\textbf{Effective oversight}
& Whether participants' evaluation successfully identified and addressed a
seeded defect.
& Binary outcome: \textit{resolved} vs.\ \textit{unresolved}.\\

\bottomrule
\end{tabularx}
\end{table}
%Our study produced multiple synchronized data streams, including interaction logs, task outcomes, self-reports, and gaze measures. Analysing these multimodal data required examining how different measures of oversight related to one another. 
% To maintain conceptual and terminological consistency %throughout the remainder of the paper, 
% we distinguish among three related but non-equivalent dimensions of oversight: perceived, behavioral, and effective oversight (Table~\ref{tab:oversight_dimensions}). We use these definitions throughout the following analysis and discussion. \textcolor{red}{change; HOw we measure and then go to the name, how users bee}

Based on how oversight appears in our data and how it can be measured, we developed three working definitions: \textit{perceived}, \textit{behavioral}, and \textit{effective oversight} (Table~\ref{tab:oversight_dimensions}). These capture, respectively, how users assess their own oversight, how oversight is shown in their behavior, and whether they actually resolved the silent defect. We use these definitions throughout the analysis and discussion.

%We organize the results around four exploratory analyses. Each subsection describes the measures and analytical procedures used, followed by the significant findings and relevant null results. We report these null results to clarify which measures did not distinguish between experimental conditions or task outcomes.

\subsection{General Findings}
\subsubsection{Sample and data quality.} The core cohort comprised 36 participants, each assigned six main tasks across two rounds: two Silent Defect (SD), two No Defect (ND), and two Blatant Defect (BD) tasks. This produced 216 task instances. Two participants lacked usable outcome data for their final task. One participant recorded no interaction with their final SD task. %, apparently because the session ended before they reached it. 
The other engaged substantially with their final BD task (10 agent turns and six pre-merge check attempts), but its outcome was not recorded. % because of a logging issue.
Both losses occurred at the sixth and final task position, suggesting end-of-session attrition rather than systematic loss under a particular condition. We retained these instances where their available data supported the analysis but excluded them from analyses requiring a final task outcome. This yielded 71 SD, 72 ND, and 71 BD tasks with usable outcomes (214 in total).

% \subsubsection{Familiarity and Task effects.} Resolution of silent defects increased from 10 of 36 tasks in Round~1 (27.8\%) to 18 of 35 in Round~2 (51.4\%). Teh effect of tasks was significant, sadly our pilot studies didnt reproduce hi behavior for us to change... We see silent resolved rates ranginfrom 8\%-25\% . but the cofindec rates or the duration spent on the tassk didnt relly depend on this. and we controlled and ranodmized the sets so any round related results remain neutral unless we account for the exact position of the task. All hereafter analysis will reflect and discuss on this.~see appendix rf table for this

\subsubsection{Task Effects}

Silent-defect resolution increased from 10/36 tasks in Round~1 (27.8\%) to 18/35 in Round~2 (51.4\%). Because task sets were randomized across rounds, this difference was not attributable to a fixed task-set assignment. However, resolution still varied substantially by individual task, ranging from 8\% for \textsc{Conflicts} to 75\% for \textsc{Billing}. We therefore account for task identity in subsequent analyses. Full task-level results are reported in Appendix Table~\ref{tab:task-difficulty-spread}.

% Evaluation also extended beyond generic checking more often in Round~2 (62.9\% vs.\ 50.0\%), and participants more frequently produced evidence capable of revealing the seeded defect (57.1\% vs.\ 41.7\%). %Although the increase in resolution approached but did not reach conventional 
% This increase was not statistically significant (Fisher's exact test, $p=.054$). % the pattern appeared across multiple stages of the evaluation process. 
% This may reflect increasing familiarity or practice. 
% % However, round effects cannot be fully separated from task-set effects because each participant received a different task set in each round.

\subsubsection{Individual differences.} Self-reported Python proficiency was associated with effective oversight. Silent-defect resolution increased from 16.7\% among beginners ($n=6$) to 39.1\% among intermediate programmers ($n=23$) and 57.1\% among advanced programmers ($n=7$; Spearman's $\rho=.34$, $p=.041$).
% Mean evaluation depth[average number of checks] followed the same pattern, although the association did not reach statistical significance ($\rho=.30$, $p=.078$). 
In contrast, self-reported experience with agentic AI tools was not associated with defect resolution ($p>.48$). Participants who frequently delegated multi-step tasks to AI were therefore no more thorough or effective than less frequent users. We also observed no association between gender and either measure (all $p>.56$; 23 men, 12 women, and one non-binary participant).

\subsection{Confidence and Perceived Thoroughness Capture Different Aspects of Oversight}
\label{sec:results-self-report}

After each main task, we asked participants to rate, on a five-point scale, (1) how confident they were that their final solution met all the requirements and (2) how thoroughly they had checked the coding agent's suggestions. Although both measures concern participants' perceptions of their own oversight, they showed notably different relationships with behavior and outcome.

\subsubsection{Confidence remained high when defects were unresolved}
Self-reported confidence did not reliably distinguish whether participants had detected a silent defect. Among the 43 failed silent-defect tasks with valid confidence ratings, 34 (79\%) received a confidence rating of 4 or 5 out of 5. Participants were therefore not simply uncertain when they handed off an incorrect solution; they were frequently highly confident that it met the requirements.

\begin{figure*}[t]
    \centering
    \includegraphics[width=0.75\textwidth]{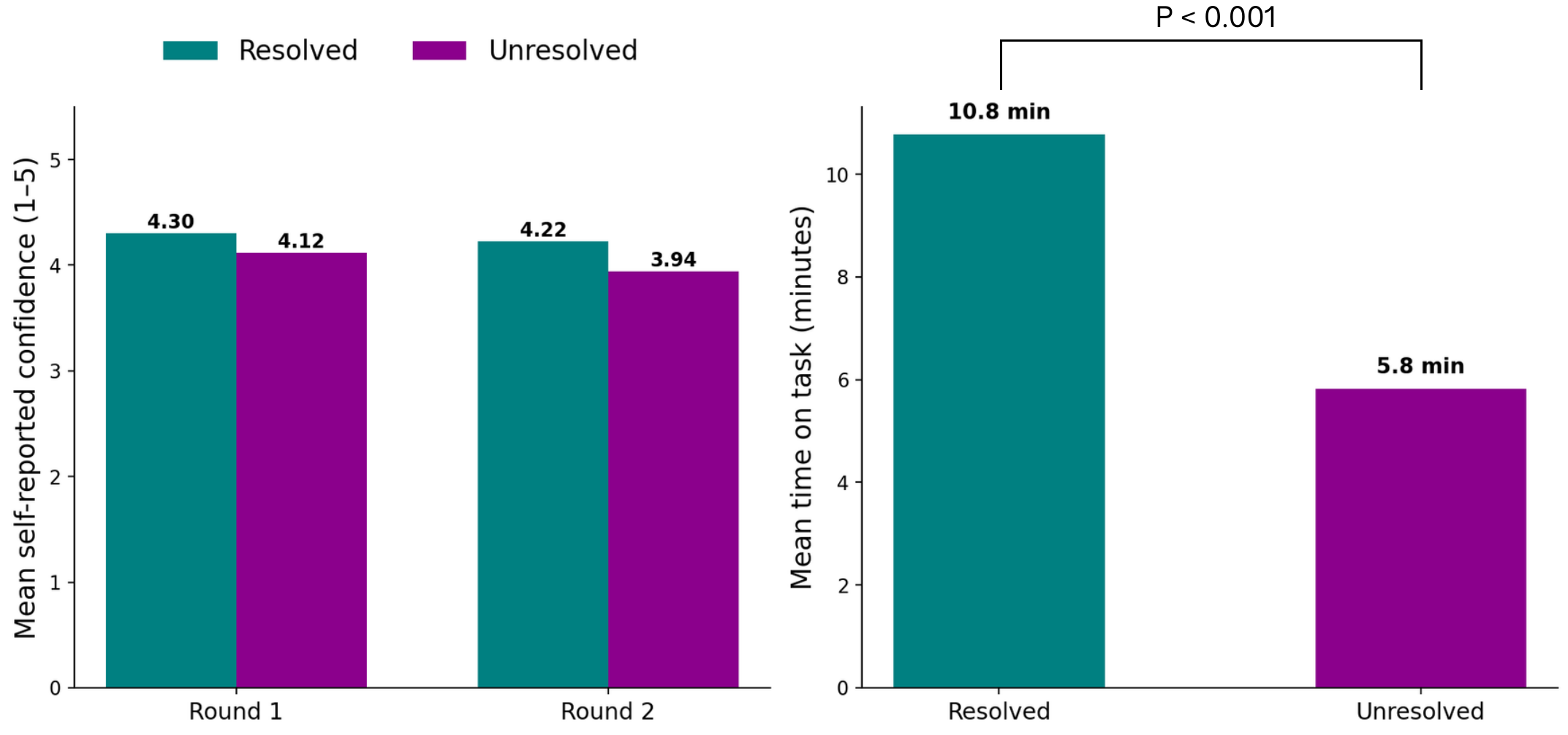}
    \caption{Confidence and task duration for resolved and unresolved silent-defect instances. Self-reported confidence remained similarly high across outcomes and rounds (left), whereas resolved defects involved substantially more time on task than unresolved defects (10.8 vs.\ 5.8 minutes, $p<.001$; right).}
    \Description{Two bar charts. The left chart plots mean self-reported confidence on a one-to-five scale for Round 1 and Round 2, with paired bars for resolved (teal) and unresolved (purple) silent defects. Round 1 values are 4.30 resolved and 4.12 unresolved; Round 2 values are 4.22 resolved and 3.94 unresolved. All four bars sit above 3.9. The right chart plots mean time on task in minutes, with a single bar per outcome: 10.8 minutes for resolved defects and 5.8 minutes for unresolved. A bracket spanning the two bars is labeled $P < 0.001$.}
    \label{fig:barchart}
\end{figure*}

Confidence remained high in both rounds. Mean confidence on silent-defect tasks decreased only slightly, from 4.17 in Round~1 to 4.09 in Round~2, and this difference was not statistically significant. Even after participants saw the broader release results for their Round~1 implementations, we found no evidence that their confidence became better calibrated in Round~2. As depicted in Figure~\ref{fig:barchart} confidence remained high for both passed and failed silent defects: 4.30 versus 4.12 in Round~1 and 4.22 versus 3.94 in Round~2, respectively. Task behavior provided a clearer distinction. Participants spent nearly twice as long on silent-defect tasks in which they recognized the defect (M = 10.8 min) than on those in which they missed it (M = 5.8 min; Mann-Whitney U, p < .001). Thus, confidence provided little indication of an impending oversight failure, even when participants' task behavior differed.

\subsubsection{Perceived thoroughness tracked within-person changes in verification}
Perceived thoroughness showed a different pattern. Across all 214 main-task observations, a participant tended to report greater thoroughness on tasks where they spent more time than was typical for them ($r_{\mathrm{within}} = 0.258$, $p < .001$). They also reported greater thoroughness when they performed more manual code executions than was typical for them ($r_{\mathrm{within}} = 0.158$, $p = .021$). In contrast, the corresponding between-person relationships were not significant. A participant who generally used higher thoroughness ratings was not necessarily more behaviorally thorough than another participant who generally used lower ratings.

This within-person relationship also extended to silent-defect outcomes. Caught silent-defect tasks received a mean thoroughness rating of 4.14, compared with 3.49 for missed tasks ($r = 0.270$, $p = .023$). A stricter within-participant comparison produced the same pattern. Among the 16 participants who caught one silent defect and missed the other, participants rated themselves an average of 0.93 points more thorough on the task in which they caught the defect ($p = .014$); 10 of the 16 participants gave a higher thoroughness rating specifically to their caught task, 4 rated themselves equally thorough on both, and 1 rated themselves less thorough on the caught task.
The changes relative to a participant's own typical rating carry meaningful information: when participants perceived themselves as being more thorough than usual, they generally spent longer, conducted more manual checks, and were more likely to detect the silent defect.

%\subsubsection*{Summary} Participants remained confident even when they missed consequential defects, and feedback on their performance after a round did not significantly improve this calibration in the following round. Perceived thoroughness, however, reflected relative changes in verification effort and defect detection within the same participant. 

\subsection{Behavior Following an External Defect Warning}
\label{sec:results-warning}
The ND, SD, and BD conditions were designed to examine how participants behaved before and after receiving an external signal that the agent could be wrong. We expected participants to approach an SD task similarly to an ND task before encountering such a signal. We further expected the failed pre-merge check and accompanying comment in the BD condition to increase scrutiny of subsequent agent output, with the strongest effect occurring on the task immediately following BD.

%Before encountering a BD warning, participants did not approach SD and ND tasks identically. They spent significantly longer on SD tasks than on ND tasks (8.68 vs.\ 6.03 minutes, $p = .004$). They also reported higher thoroughness on SD tasks (3.81 vs.\ 3.38), although this difference did not reach statistical significance ($p = .083$). We found no corresponding differences in test-run count, manual-code-run count, or confidence.

At an aggregate level, encountering a BD warning did not improve subsequent silent-defect detection. Participants caught 38.89\% of silent defects when SD occurred before BD and 40\% when SD occurred after BD. These catch rates were statistically indistinguishable ($\chi^2 = 0.00$, $p = 1.00$). Task duration also did not differ significantly depending on whether SD occurred before or after BD ($p = .170$). However, because different tasks appeared in different sequence positions and varied substantially in difficulty, the aggregate before-versus-after comparison cannot cleanly separate the effect of prior BD exposure from the effect of task identity.

However, the overall before-and-after comparison masked differences in the timing of the subsequent silent-defect task. When a silent defect immediately followed a visible failure (\textit{BD~$\rightarrow$~SD}), participants detected 50\% of defects (12/24 tasks; 12 participants). When one correct task intervened (\textit{BD~$\rightarrow$~ND~$\rightarrow$~SD}), detection fell to 9.1\% (1/11 tasks; six participants). The difference was supported by an exact participant-level permutation test (p=.046). A task-level Fisher’s exact test produced a similar result (p=.023); we treat this only as a sensitivity analysis because multiple tasks were contributed by the same participants. Because task identity differed across sequence positions, we additionally compared performance after residualizing each task by its own mean resolution rate. The adjacent-versus-gap difference remained significant after this adjustment ($t=2.44$, $p=.020$). This comparison is supported by within-task overlap, as \textsc{Equipment} and \textsc{Waitlist} appeared in both sequence groups.

% This pattern provides preliminary evidence of short-lived re-engagement. A visible failure was associated with substantially better detection when the silent defect appeared immediately afterward, but this advantage diminished after one correct interaction. The difference was not explained by greater time investment: adjacent tasks lasted 7.35 minutes on average, compared with 5.76 minutes for gap tasks, but this difference was not significant in either the participant-level permutation test (($\Delta=1.72$) minutes, (p=.494)) or a task-level Welch test ((p=.454)). Given the small sequence groups—particularly the six participants in the gap condition—and the participant-level result’s proximity to the conventional significance threshold, we interpret this as suggestive rather than conclusive evidence that failure-triggered re-engagement may decay rapidly.

%\subsubsection*{Summary.} We see preliminary evidence of short-lived re-engagement. Participants were more likely to detect a silent defect when it immediately followed a visible failure, but this advantage diminished when a correct interaction occurred between them. Given the small sequence groups, particularly the six participants in the gap condition, we interpret this as suggestive rather than conclusive evidence that failure-triggered re-engagement may decay rapidly.

\subsection%{Multimodal Signals} 
{Input inactivity reflected effective oversight}
\label{sec:multimodal}
%We examined whether input behavior, gaze dynamics, and physiological signals differed across defect conditions or were associated with effective oversight. Full feature definitions, preprocessing procedures, statistical models, and results are provided in the supplementary materials.

%\subsubsection{Input inactivity reflected effective oversight}
Across 21 keyboard and cursor measures, silent-defect and no-defect tasks differed in completion time but not in how much participants typed or moved after accounting for task duration. However, silent defects that were resolved contained a substantially greater proportion of time without typing than those that were missed (.622 vs.\ .425, $p<.001$, Cohen's $d=1.29$). 
% This pattern also appeared within the 14 participants who resolved one silent defect but missed the other (.626 vs.\ .466, $p=.016$). 
Longer gaps without keystrokes and greater variability in typing rhythm followed the same descriptive pattern, whereas keystroke rate, deletions, navigation inputs, cursor travel, and cursor speed did not distinguish the outcomes. Although periods without input are compatible with participants pausing to read the implementation, revisit the requirements, or reason about possible edge cases. Evaluation effort may therefore be more visible when interaction stops than in the amount of observable activity.

\subsubsection{Gaze dynamics did not provide a reliable task-level marker.}
We analysed 71 silent-defect tasks from 36 participants, comprising 28 resolved
and 43 unresolved defects. In pooled comparisons, resolved tasks had fewer
fixations per minute (146.97 vs.\ 167.36, $p=.042$), fewer saccades per minute
(169.04 vs.\ 196.66, $p=.034$), and smaller mean saccade amplitudes
($7.14^{\circ}$ vs.\ $8.42^{\circ}$, $p=.009$). However, these patterns did not
consistently persist within participants. Among the 16 participants who resolved
one silent defect but missed another, saccade rate and amplitude no longer
differed between outcomes (all $p\geq.064$), while fixation rate reversed
direction: participants produced 7.39 more fixations per minute during resolved
tasks ($p=.008$).

After accounting for task duration, round, task position, and repeated
observations, the estimated within-participant association between fixation rate
and defect resolution remained positive but uncertain (OR $=1.67$, 95\% CrI
$[0.87,3.22]$). In contrast, task duration showed a stronger association with
resolution (OR $=5.32$, 95\% CrI $[2.50,11.35]$). General gaze dynamics may
therefore contain some information about changes in an individual's behavior,
but they did not provide a reliable standalone marker of effective oversight.

\subsubsection{Physiological signals did not distinguish conditions or outcomes.}
The PPG signal captured by the Aria Gen~2 glasses was too noisy for reliable task-level analysis and was therefore excluded. We instead analysed cardiac and respiratory measures recorded by the Garmin watch wearable. None reliably differed across the ND, SD, and BD conditions or between resolved and unresolved silent defects. However, physiological arousal declined gradually over the session, suggesting that these measures were more sensitive to participants settling into the study environment than to the evaluation demands of individual tasks.

Full feature definitions, preprocessing procedures, statistical models, and results are provided in the supplementary materials.

\subsection{Understanding Behavioral Oversight in Agentic Interactions}
\label{sec:oversight-analysis}

% We used \cite{fern2007mining} s way of clustering to cluster patterns from our log data.

% \begin{table}[t]
% \centering
% \caption{Three related but non-equivalent aspects of oversight examined in our study.}
% \label{tab:oversight_dimensions}
% \begin{tabularx}{\columnwidth}{@{}p{0.21\columnwidth}XX@{}}
% \toprule
% \textbf{Aspect} & \textbf{Definition} & \textbf{Operationalization} \\
% \midrule
% \textbf{Perceived oversight}
% & How participants believed they had evaluated the agent's work.
% & Self-reported confidence and perceived thoroughness. \\

% \textbf{Behavioral oversight}
% & Observable actions through which participants sought and evaluated evidence beyond the agent's initial output.
% & Code inspection, requirement revisiting, manual execution, additional tests, questions to the agent, and evaluation--revision loops. \\

% \textbf{Effective oversight}
% & Whether participants' evaluation successfully identified and addressed a seeded defect.
% & Defect recognition, corrective intervention, and verification that the defect was resolved. \\
% \bottomrule
% \end{tabularx}
% \end{table}

% \textbf{Coding Behavioral Oversight}

% \subsubsection{Coding Behavioral Oversight}

% \subsubsection{Coding Behavioral Oversight}

Our analysis was informed by prior work on interaction-strategy mining~\cite{fern2007mining} and taxonomies of AI-assisted programming~\cite{mozannar2024reading}. We manually coded 216 participant-authored messages to the agent, as well as code entered in the manual execution panel, to identify participants' interaction and verification strategies. Message codes captured requests for implementation, verification, explanation, diagnosis, and correction, while scratch code indicated attempts to test the implementation with participant-generated inputs.
The first author developed the codebook using a diverse pilot sample and then coded the complete dataset. A second author independently coded a stratified 24\% sample, achieving an agreement of 75\% (linear-weighted Cohen's $\kappa$ = 0.632). Disagreements were discussed and resolved. When the meaning of a message or action was unclear, we examined the corresponding first-person video segment for context. We then combined these manually coded data with automatically recorded actions, including task-brief revisits, manual program executions, pre-merge checks, code edits, and task completion. This produced a chronological record of oversight behavior for each participant--task pair.

\begin{figure*}[t]
    \centering
    \includegraphics[width=\textwidth]{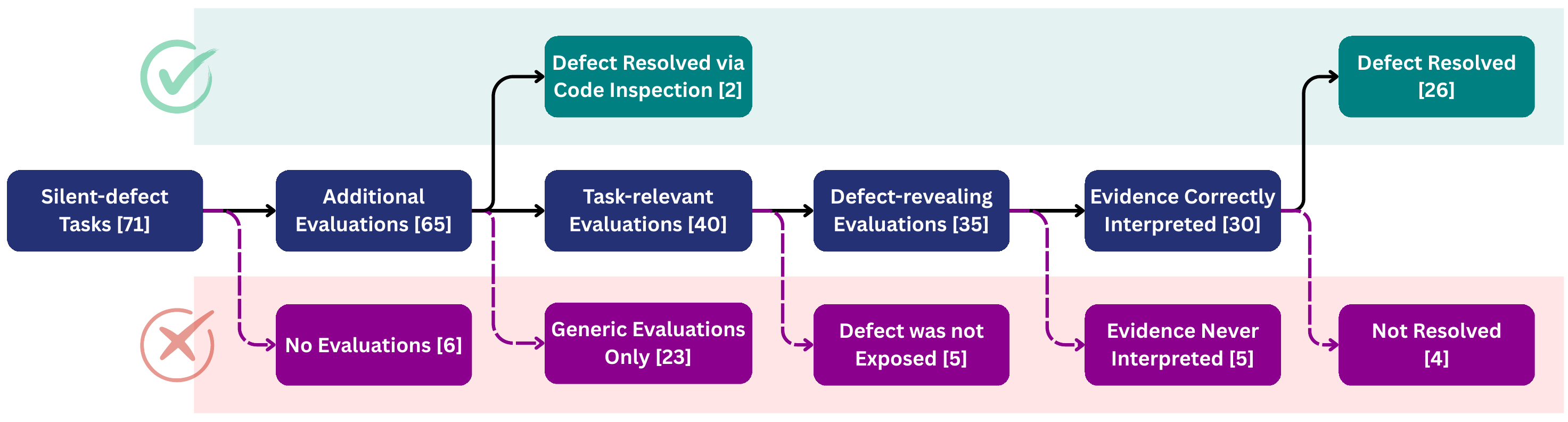}
    \caption{Progression of the 71 silent-defect task instances through the oversight pipeline. The main pathway shows task-relevant evaluation, production of defect-revealing evidence, correct interpretation, and resolution. Purple branches indicate where task instances stopped progressing. Two additional defects were resolved through non-targeted evaluation: one through generic verification and one through direct code inspection.}
    \Description{A left-to-right flow diagram tracing 71 silent-defect task instances through five stages. The main pathway, marked with a green check, runs along the top: Silent-defect Tasks [71], Additional Evaluations [65], Task-relevant Evaluations [40], Defect-revealing Evaluations [35], Evidence Correctly Interpreted [30], and Defect Resolved [26]. A parallel row below, marked with a red cross, shows where instances stopped progressing: No Evaluations [6], Generic Evaluations Only [23], Defect was not Exposed [5], Evidence Never Interpreted [5], and Not Resolved [4]. A separate box, Defect Resolved via Code Inspection [2], branches upward out of the main sequence before the task-relevant stage.}
    \label{fig:oversight-pipeline}
\end{figure*}

\subsubsection{Where Oversight Breaks Down}
To characterize how participants evaluated the agent's output before completing a silent-defect task, we traced all 71 task instances through these stages (Figure~\ref{fig:oversight-pipeline}). To illustrate, consider the \textsc{Billing} task, in which participants asked the agent to reconcile a member's charges into a running total, distinguishing between approved, refunded, and cancelled amounts. The seeded defect ignored refunds instead of subtracting them. Generic evaluation, such as rerunning the provided example or asking the agent to check its code, would not necessarily expose this defect. Even a task-relevant test involving cancelled charges would not reveal it. In contrast, testing an approved charge together with an equal refund would expose the error because the balance should be zero. Effective oversight required participants to notice this mismatch, correct the refund logic, and verify the revised solution. The defect could also be identified directly by inspecting the code.

Most instances (65/71, 91.5\%) included at least one evaluative action beyond the mandatory pre-merge check, whereas six relied on that check alone. The relevance of these additional actions varied substantially. In 23 instances, evaluation remained generic, such as rerunning an existing example or asking the agent to "check" the solution without specifying what to examine. In 40 instances, participants evaluated risks associated with the task requirements. Of these, 35 produced evidence capable of revealing the seeded
defect, whereas five examined plausible but unrelated risks.

Even revealing evidence did not guarantee successful oversight. In five of the 35 instances, participants produced evidence exposing the defect but did not recognize its implication. Participants recognized the defect in the remaining 30 instances, although four did not successfully resolve it. Thus, 26 defects were resolved through task-relevant evaluation. Two remaining defects were resolved through manual code inspection and editing.

\subsubsection{Oversight Behavior Quality improved in Round 2}
\begin{table}[t]
\centering
\caption{Progress through the oversight pipeline across study rounds. Percentages are calculated relative to the number of silent-defect task instances in each round.}
\label{tab:pipeline-rounds}
\small
\begin{tabular}{lcc}
\toprule
\textbf{Pipeline stage} & \textbf{Round 1} ($n=36$) & \textbf{Round 2} ($n=35$) \\
\midrule
Additional evaluation
    & 33 (91.7\%) & 32 (91.4\%) \\
Task-relevant evaluation
    & 18 (50.0\%) & 22 (62.9\%) \\
Defect-revealing evidence
    & 15 (41.7\%) & 20 (57.1\%) \\
Defect recognized
    & 11 (30.6\%) & 19 (54.3\%) \\
Resolved through main pathway
    & 9 (25.0\%) & 17 (48.6\%) \\
Total resolved
    & 10 (27.8\%) & 18 (51.4\%) \\
\bottomrule
\end{tabular}
\end{table}

Table~\ref{tab:pipeline-rounds} compares participants' progression through the oversight pipeline across rounds. Participants were similarly likely to initiate additional evaluation in Round~1 and Round~2 (91.7\% vs.\ 91.4\%). However, their evaluations progressed further in Round~2: more became task-relevant (62.9\% vs.\ 50.0\%), produced defect-revealing evidence (57.1\% vs.\ 41.7\%), and resulted in correct defect recognition (54.3\% vs.\ 30.6\%). Correspondingly, the overall resolution rate increased from 27.8\% in Round~1 to 51.4\% in Round~2. This pattern suggests that the improvement was not driven by participants checking more often, but by their checks becoming better targeted and more effectively interpreted. To test these round differences, we fit separate logistic regression models for each binary pipeline outcome, with round and task identity as predictors and participant-clustered standard errors to account for repeated observations. The round effect remained significant for defect recognition ($p=.005$) and resolution ($p=.009$), was not significant for defect-revealing evidence ($p=.078$), and task-relevant evaluation ($p=.170$).

\begin{figure*}[t]
    \centering
    \includegraphics[width=\textwidth]{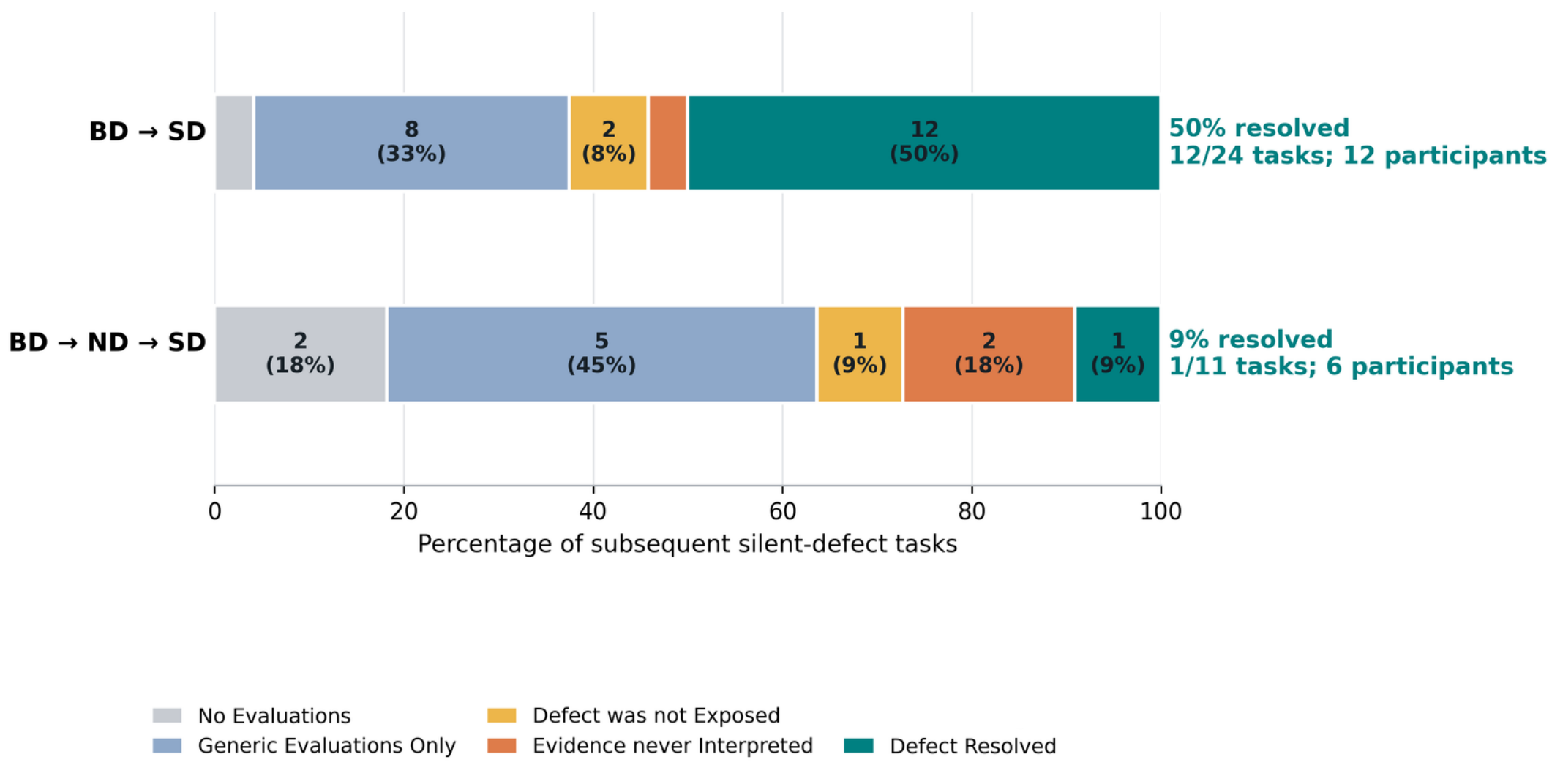}
    \caption{Oversight outcomes on silent-defect tasks following a visible failure. When the silent-defect task immediately followed the visible failure, participants resolved 12 of 24 defects (50\%). When one correct task intervened, they resolved only 1 of 11 defects (9\%). Bars show the percentage of tasks ending at each stage of the verification pathway; values inside the bars show task counts and within-sequence percentages.}
    \Description{A horizontal stacked bar chart with two rows, showing the percentage of subsequent silent-defect tasks ending at each stage of the verification pathway. The BD $\rightarrow$ SD row divides into generic evaluations only, 8 tasks (33\%); defect was not exposed, 2 tasks (8\%); evidence never interpreted, 2 tasks (8\%); and defect resolved, 12 tasks (50\%). It is annotated 50\% resolved, 12 of 24 tasks, 12 participants. The BD $\rightarrow$ ND $\rightarrow$ SD row divides into no evaluations, 2 tasks (18\%); generic evaluations only, 5 tasks (45\%); defect was not exposed, 1 task (9\%); evidence never interpreted, 2 tasks (18\%); and defect resolved, 1 task (9\%). It is annotated 9\% resolved, 1 of 11 tasks, 6 participants. The resolved segment is far longer in the first row than the second.}
    \label{fig:visible-failure-carryover}
\end{figure*}

\subsubsection{Re-engagement after visible failure was short-lived.}
Figure~\ref{fig:visible-failure-carryover} shows how participants evaluated the next silent-defect task after encountering a visible failure. When the silent defect appeared immediately afterward, participants resolved 12 of 24 instances (50\%). In contrast, when one correct task occurred between the visible failure and the silent defect, only 1 of 11 instances (9\%) was resolved. The difference was supported by an exact participant-level permutation test ($p=.046$). The pipeline endpoints further suggest that evaluation was more likely to remain generic or stop before correctly interpreting the evidence after an intervening correct task. This pattern is consistent with visible failure temporarily reactivating effective oversight, with the effect weakening after a single successful interaction (Section~\ref{sec:results-warning}).

\subsection{Summary}
%Taken together, our analyses show that effective oversight could not be inferred from any single measure. Participants often completed a task soon after the provided checks passed, yet passing these checks did not guarantee that the solution was correct. Additional evaluation was also not necessarily effective: participants sometimes performed generic checks, examined irrelevant risks, or failed to interpret evidence that exposed the defect. As a whole, effective oversight depended on moving beyond the system's default evidence of success to generate, interpret, and act on evidence grounded in the task requirements.
%
%Self-reports and multimodal signals captured different parts of this process. Confidence remained high even when silent defects were missed, while perceived thoroughness was informative mainly relative to a participant's own baseline. Longer task duration and periods without typing were associated with successful defect resolution, but gaze and physiological measures did not provide reliable standalone markers. %Across modalities, differences between participants were generally larger than differences between effective and ineffective oversight within the same participant.

Participants remained confident even when they missed consequential defects, and feedback on their performance after a round did not significantly improve this calibration in the following round. Perceived thoroughness, however, reflected relative changes in verification effort and defect detection within the same participant. 

Self-reports and multimodal signals captured different parts of this process. Confidence remained high even when silent defects were missed, while perceived thoroughness was informative mainly relative to a participant's own baseline. Longer task duration and periods without typing were associated with successful defect resolution, but gaze and physiological measures did not provide reliable standalone markers.

Also, we see preliminary evidence of short-lived re-engagement. Participants were more likely to detect a silent defect when it immediately followed a visible failure, but this advantage diminished when a correct interaction occurred between them. Given the small sequence groups, particularly the six participants in the gap condition, we interpret this as suggestive rather than conclusive evidence that failure-triggered re-engagement may decay rapidly.

% These exploratory findings do not establish a sensor-based detector of mindlessness. Instead, they characterize oversight as a contextual process in which observable activity, perceived effort, defect recognition, and final correctness may diverge. Among the measures examined, task-grounded interaction sequences provided the clearest account of whether participants moved beyond surface evidence and meaningfully evaluated the agent's work.

% \section{Meta Cognitive Nudge}
% \input{Files/05-MetaCognitive_Nudge}

\section{Discussion}

Across our investigation, we found that effective oversight can break down even when people appear actively engaged with an agent's work. Participants frequently expressed high confidence after missing a silent defect, and this confidence did not recalibrate after they had seen the outcomes of the previous round. Behaviorally, participants who detected a defect tended to spend longer on the task and showed more pauses without typing. Physiological measures, however, showed substantial between-participant variation and little consistent within-participant differentiation, limiting their usefulness as general indicators of oversight while suggesting that any sensing-based intervention may need to be personalized. We also found suggestive evidence that a visible failure could briefly prompt closer scrutiny of the next task, although this effect appeared to fade quickly. In the sections that follow, we discuss what these patterns mean for cognitive engagement in AI-assisted work, how they relate to authorship and responsibility as agents take on more of the task, and how failure cues may shape the timing of effective oversight.

\subsection{Cognitive Engagement and Delegation}
Cognitive engagement cannot be inferred simply from whether users remain active. Participants inspected code, ran tests, and sent follow-up prompts, but these actions alone did not show whether they had evaluated the assumptions on which correctness depended. Therefore, we coded participants' prompts, scratch runs, tests, and inspection behavior according to whether they produced evidence relevant to the seeded defect. This coding does not capture every internal judgment, but it distinguishes generic activity from observable, task-relevant evaluation. 
As shown in Figure~\ref{fig:oversight-pipeline}, effective oversight required participants to identify what could fail, produce evidence capable of exposing it, and interpret that evidence correctly.

The interviews help explain these behavioral differences. P29 relied on the available checks: \emph{"I ran the test cases and pre-merge check, and when both passed, I assumed the solution was correct."} Others delegated parts of the evaluation to the agent. P33 reported, \emph{"I mostly copied the instructions and asked it to handle edge cases,"} while P27 asked the agent to generate tests and then ran them. These actions showed that participants attempted verification, but not necessarily that they understood what risks were being tested or what remained unchecked.

By contrast, P26 noticed that an important requirement had not been exercised: \emph{"The supplied example did not test case-insensitive behavior. When I tested that requirement myself, it failed."} This required recognizing the limits of the available evidence and constructing a diagnostic example. Yet generating relevant evidence was still not sufficient in every case. Some participants produced outputs that exposed a defect but did not interpret them correctly.

These findings extend concerns about cognitive offloading in AI-supported work~\cite{gerlich2025ai,lee2025impact}. Agentic systems allow users to delegate implementation, but users may also delegate the reasoning needed to evaluate it. Asking the same agent to \emph{"check the code"} can create visible verification without an independent judgment about where the implementation may fail. Delegating test generation is not inherently ineffective. Its value depends on whether users understand what was tested, recognize what remains untested, and can interpret the results.

\subsection{Authorship and Responsibility}
Agentic programming creates a mismatch between authorship and responsibility: users remain responsible for evaluating code they did not write or closely follow. Our formative interviews anticipated this tension. F4 explained, \emph{"If we are not acting on something, and we are letting something else act on our behalf, then it can be adverse for us in terms of being responsible for our actions."} Delegating execution does not remove responsibility for the result, but it may reduce the understanding needed to evaluate it.

The post-study interviews show how this mismatch becomes more difficult when generated code exceeds the user's knowledge. P29 explained, \emph{"I understood the expected output but not all of the implementation, so I accepted the code and ran the test cases."} P39 similarly accepted generated code because reproducing the work independently would take too long. Consistent with these accounts, programming experience was associated with silent-defect detection, yet less-experienced participants were not correspondingly less confident. Coding agents may therefore enable users to produce implementations beyond what they can independently evaluate, while leaving them responsible for the outcomes.

The amount of responsibility users assumed also depended on the perceived stakes of the task. In both studies, participants described relying on final outputs, accepting generated work under time pressure, and intentionally delegating thought during familiar or low-stakes tasks. P24 stated, \emph{"It depends on how serious the task is. For a high-stakes task, I would check more. These tasks felt low stakes, so I did not feel the need to put in as much effort."} Such delegation may be reasonable when the consequences are limited, as time and effort are saved. The risk arises when users rely on cues such as plausible code and passing checks to decide that deeper evaluation is unnecessary, even when those cues provide incomplete evidence. This reflects Langer's account of mindlessness: familiar cues guide behavior while distinctions that matter in the current situation are overlooked~\cite{langer1978mindlessness,langer1989minding}.

Our results suggest that simply leaving final approval to the user is therefore not sufficient to establish meaningful human responsibility. Participants often remained procedurally involved, yet comparatively few evaluations progressed far enough to solve and correctly interpret the silent defect. In such cases, users may retain formal responsibility without having developed the task understanding needed to exercise it effectively. For agentic systems, supporting responsibility may therefore require more than preserving a final confirmation step: systems should help users maintain enough task-relevant understanding to evaluate consequential decisions, particularly when generated work exceeds their own expertise.

%We therefore distinguish three aspects of oversight: \emph{perceived oversight}, how vigilant users believe they were; \emph{behavioral oversight}, the actions through which they seek and evaluate evidence; and \emph{oversight effectiveness}, whether that process identifies and addresses the defect. These dimensions did not always align. Neither subjective vigilance, visible activity, nor final correctness is sufficient on its own to characterize oversight.

\subsection{Failure Cues and Effective Oversight}
Visible failure gave users a clear reason to question the agent. Several participants reported checking later tasks more carefully after encountering or learning about a failure. P42 explained, \emph{"I failed one of the three tasks even though I thought I had done well on all three. That made me think there might be a trap in the questions, so I checked more thoroughly in the second round."} P28 similarly reported treating later outputs as potentially suspicious after noticing an obvious error. These accounts support our behavioral evidence that failure cues can interrupt routine acceptance and prompt re-engagement.

However, this effect appeared short-lived. Silent defects were more often detected when they immediately followed a visible failure, but this advantage diminished when a correct interaction occurred between them. Apparent success may therefore restore trust and again reduce the perceived need for scrutiny. This resembles established patterns of automation complacency and over-reliance: reliable performance reduces monitoring, while apparent success encourages users to defer to the system rather than seek independent evidence~\cite{bainbridge1983ironies,parasuraman1993performance,parasuraman2010complacency,skitka2000automation}.

As discussed earlier, the depth of checking depended on its perceived value. Participants reduced oversight when tasks were low stakes, time was limited, or the generated code was too large or unfamiliar to inspect efficiently. Both studies also raised concerns about learning, critical thinking, and the gradual loss of opportunities to reason through problems independently. P31 distinguished completion from understanding: \emph{"AI may help me finish that work, but whether it helps me learn is a different question. Often, it does not."} Participants therefore valued deeper involvement in learning and research contexts, while accepting greater delegation when obtaining a working result was the main goal.

This creates an important tension. Delegation can reduce immediate effort, but effective oversight still depends on the knowledge and reasoning needed to evaluate the resulting work. Effective oversight should not require constant vigilance, as delegating routine work can be appropriate. The problem is miscalibration: users may not recognize when limited checking is insufficient or when passing checks provide weak evidence of correctness. This was visible in the gap between confidence, observable evaluation, and defect detection. The challenge may be particularly important for less-experienced users, as coding agents can enable them to produce implementations that exceed their ability to evaluate independently. %In our study, programming experience was associated with defect detection, yet less-experienced participants were not correspondingly less confident. Even after receiving performance feedback and encountering failures, some participants continued to overestimate the adequacy of their oversight.

\subsection{Limitations and Future Work}

Behavioral oversight is not a direct measure of internal cognitive state. Interaction logs capture observable actions, but limited activity may reflect active reasoning, while frequent activity may remain superficial. Our inactivity-related findings reinforce the importance of considering the timing and context of behavior rather than activity alone. We therefore characterize observable oversight rather than claim direct detection of reduced oversight or mindlessness. Future work should validate behavioral indicators against independently coded, task-relevant reasoning at finer temporal resolution.

We acknowledge several limitations in the study design. First, we used short Python tasks, seeded defects, and a controlled coding agent. This ensured comparable exposure to correct, silently defective, and visibly failing outputs, but did not capture the scale or uncertainty of real development and today's capable AI agents. This may have affected how participants inspected and corrected the code, differing from their natural behavior. Future studies should examine naturally occurring failures in larger repositories using current coding agents. Second, silent-defect detectability varied substantially by task, from 8\% for CONFLICTS to 75\% for BILLING. This difference did not emerge during piloting. Although task assignment and order were randomized, the aggregate findings remain specific to the defects sampled.

% Detectability also varied substantially across tasks. Silent-defect resolution ranged from 8\% for CONFLICTS to 75\% for BILLING. Randomizing assignment and order reduced systematic confounding with participants and sequence position, and our pilots did not reveal similarly large differences. However, randomization does not make these defects representative of silent failures more broadly. Aggregate rates should therefore be interpreted in relation to the sampled tasks. Future studies should systematically vary requirement salience, semantic subtlety, and testability across a larger task set.

% Participants knew that their work was being evaluated, and first-round feedback explicitly encouraged closer checking. They may therefore have exercised more oversight than during everyday use. Longitudinal studies are needed to examine how oversight changes as users develop familiarity and trust in agents over time.

Several exploratory analyses may have been underpowered. In particular, the visible-failure carryover comparison included only six participants in the gap sequence, and changes across rounds may also reflect learning, task identity, or position. The findings suggest that failure can prompt re-engagement, but do not establish how reliable or lasting this effect is. Larger and longitudinal studies should independently manipulate warning timing and examine how oversight changes as users develop familiarity and trust over time.

Our study did not evaluate specific interventions, but it identifies directions for task-calibrated re-engagement. Future systems could clarify what passing checks establish, highlight untested requirements, and support users in constructing and interpreting task-relevant evidence. Such support could be explored after failures, at consequential handoffs, or when existing checks provide limited coverage, while accounting for task stakes, familiarity, and reversibility. Future studies should determine whether these approaches improve oversight, for whom, and at what cost to workload and the benefits of delegation.

Finally, the sensing analyses were limited by measurement quality and granularity. Whole-task physiological summaries may obscure brief changes in engagement. Clear differences between participants also suggest that universal thresholds may be unsuitable. Our findings do not establish reliable real-time detection, but motivate event-aligned, person-specific models validated against independently coded oversight.

\section{Conclusion}
When AI agents execute more of the work, the human role shifts to overseeing the work. Our results show that this oversight fails in a specific, measurable way. Across 36 participants, oversight broke down not because people stopped acting, but because their actions didn't amount to the required evaluation. Participants ran tests, inspected outputs, but never produced the one piece of evidence that would have exposed a silent defect. Sometimes, they even misinterpreted the generated evidence. %Passing checks functioned as stopping cues at nearly identical rates whether the code was correct or secretly broken. 
Confidence was flat across tasks, even though defect detection varied substantially across tasks. In short, the code looked fine to them, leading participants to believe their evaluation was sufficient when it was not.

By separating perceived, behavioral, and effective oversight, we show that no single proxy-activity, confidence, or final correctness captures whether a person is truly overseeing an agent. This does not imply users need constant vigilance; exhaustive review would forfeit the point of delegation. It is that oversight must be \textit{calibrated}: users need help recognizing when the evidence in front of them cannot support the conclusion they are about to draw. Visible failures could prompt this re-engagement, but the effect was not always sustained. Through this study, we highlight a central challenge for agentic interaction: understanding when deeper oversight is needed. We identify where oversight breaks down and show how task context, interaction behavior, and person-specific patterns provide a foundation for addressing this challenge.

% %% The next two lines define the bibliography style to be used, and
% %% the bibliography file.
\bibliographystyle{ACM-Reference-Format}
\bibliography{sample-base}

@article{gerlich2025ai,
  title={AI tools in society: Impacts on cognitive offloading and the future of critical thinking},
  author={Gerlich, Michael},
  journal={Societies},
  volume={15},
  number={1},
  pages={6},
  year={2025},
  publisher={Multidisciplinary Digital Publishing Institute}
}

@inproceedings{lee2025impact,
  title={The impact of generative AI on critical thinking: Self-reported reductions in cognitive effort and confidence effects from a survey of knowledge workers},
  author={Lee, Hao-Ping and Sarkar, Advait and Tankelevitch, Lev and Drosos, Ian and Rintel, Sean and Banks, Richard and Wilson, Nicholas},
  booktitle={Proceedings of the 2025 CHI conference on human factors in computing systems},
  pages={1--22},
  year={2025}
}

@article{buccinca2021trust,
  title={To trust or to think: cognitive forcing functions can reduce overreliance on AI in AI-assisted decision-making},
  author={Bu{\c{c}}inca, Zana and Malaya, Maja Barbara and Gajos, Krzysztof Z},
  journal={Proceedings of the ACM on Human-computer Interaction},
  volume={5},
  number={CSCW1},
  pages={1--21},
  year={2021},
  publisher={ACM New York, NY, USA}
}

@inproceedings{chen2026code,
  title={Code with me or for me? how increasing ai automation transforms developer workflows},
  author={Chen, Valerie and Talwalkar, Ameet and Brennan, Robert and Neubig, Graham},
  booktitle={Proceedings of the 2026 CHI Conference on Human Factors in Computing Systems},
  pages={1--19},
  year={2026}
}

@inproceedings{mccabe2024ironies,
  title={Ironies of programming automation: Exploring the experience of code synthesis via large language models},
  author={McCabe, Alan T and Bj{\"o}rkman, Moa and Engstr{\"o}m, Joel and Kuang, Peng and S{\"o}derberg, Emma and Church, Luke},
  booktitle={Companion Proceedings of the 8th International Conference on the Art, Science, and Engineering of Programming},
  pages={12--21},
  year={2024}
}

@incollection{bainbridge1983ironies,
  title={Ironies of automation},
  author={Bainbridge, Lisanne},
  booktitle={Analysis, design and evaluation of man--machine systems},
  pages={129--135},
  year={1983},
  publisher={Elsevier}
}

@article{endsley1995out,
  title={The out-of-the-loop performance problem and level of control in automation},
  author={Endsley, Mica R and Kiris, Esin O},
  journal={Human factors},
  volume={37},
  number={2},
  pages={381--394},
  year={1995},
  publisher={SAGE Publications Sage CA: Los Angeles, CA}
}

@article{parasuraman1993performance,
  title={Performance consequences of automation-induced'complacency'},
  author={Parasuraman, Raja and Molloy, Robert and Singh, Indramani L},
  journal={The international journal of aviation psychology},
  volume={3},
  number={1},
  pages={1--23},
  year={1993},
  publisher={Taylor \& Francis}
}

@article{langer1978mindlessness,
  title={The mindlessness of ostensibly thoughtful action: The role of" placebic" information in interpersonal interaction.},
  author={Langer, Ellen J and Blank, Arthur and Chanowitz, Benzion},
  journal={Journal of personality and social psychology},
  volume={36},
  number={6},
  pages={635},
  year={1978},
  publisher={American Psychological Association}
}

@incollection{langer1989minding,
  title={Minding matters: The consequences of mindlessness--mindfulness},
  author={Langer, Ellen J},
  booktitle={Advances in experimental social psychology},
  volume={22},
  pages={137--173},
  year={1989},
  publisher={Elsevier}
}

@inproceedings{fan2026help,
  title={When Help Hurts: Verification Load and Fatigue with AI Coding Assistants},
  author={Fan, Guangrui and Liu, Dandan and Pan, Lihu and Zhang, Rui},
  booktitle={Proceedings of the 2026 CHI Conference on Human Factors in Computing Systems},
  pages={1--25},
  year={2026}
}

@inproceedings{kazemitabaar2025exploring,
  title={Exploring the design space of cognitive engagement techniques with ai-generated code for enhanced learning},
  author={Kazemitabaar, Majeed and Huang, Oliver and Suh, Sangho and Henley, Austin Z and Grossman, Tovi},
  booktitle={Proceedings of the 30th international conference on intelligent user interfaces},
  pages={695--714},
  year={2025}
}

@article{tang2026coding,
  title={How coding agents fail their users: A large-scale analysis of developer-agent misalignment in 20,574 real-world sessions},
  author={Tang, Ningzhi and Chen, Chaoran and Xu, Gelei and Shi, Yiyu and Huang, Yu and McMillan, Collin and Dong, Tao and Li, Toby Jia-Jun},
  journal={arXiv preprint arXiv:2605.29442},
  year={2026}
}

@inproceedings{hutchison2026tab,
  title={To Tab or Not to Tab: Measuring Critical Engagement in AI Code Completion Tools Using Behavioral Signals and Attention Checks},
  author={Hutchison, Jessica and Applebaum, Ian Tyler and Angelikas, Kenneth and Patel, Kush Rakesh and Nguyen, Phuoc and Lazaro, Antonio and Rucinski, Nicholas and Nabid, Rahad Arman and MacNeil, Stephen},
  booktitle={Proceedings of the 31st ACM Conference on Innovation and Technology in Computer Science Education V. 1},
  pages={217--223},
  year={2026}
}

@inproceedings{mozannar2024reading,
  title={Reading between the lines: Modeling user behavior and costs in AI-assisted programming},
  author={Mozannar, Hussein and Bansal, Gagan and Fourney, Adam and Horvitz, Eric},
  booktitle={Proceedings of the 2024 CHI conference on human factors in computing systems},
  pages={1--16},
  year={2024}
}

@inproceedings{fern2007mining,
  title={Mining interpretable human strategies: A case study},
  author={Fern, Xiaoli Z and Komireddy, Chaitanya and Burnett, Margaret},
  booktitle={Seventh IEEE International Conference on Data Mining (ICDM 2007)},
  pages={475--480},
  year={2007},
  organization={IEEE}
}

@article{parasuraman1997humans,
  author  = {Raja Parasuraman and Victor Riley},
  title   = {Humans and Automation: Use, Misuse, Disuse, Abuse},
  journal = {Human Factors},
  volume  = {39},
  number  = {2},
  pages   = {230--253},
  year    = {1997},
  doi     = {10.1518/001872097778543886}
}

@article{skitka2000automation,
  author  = {Linda J. Skitka and Kathleen L. Mosier and Mark Burdick and Bonnie Rosenblatt},
  title   = {Automation Bias and Errors: Are Crews Better Than Individuals?},
  journal = {The International Journal of Aviation Psychology},
  volume  = {10},
  number  = {1},
  pages   = {85--97},
  year    = {2000},
  doi     = {10.1207/S15327108IJAP1001_5}
}

@article{dzindolet2003trust,
  author  = {Mary T. Dzindolet and Scott A. Peterson and Regina A. Pomranky and Linda G. Pierce and Hall P. Beck},
  title   = {The Role of Trust in Automation Reliance},
  journal = {International Journal of Human-Computer Studies},
  volume  = {58},
  number  = {6},
  pages   = {697--718},
  year    = {2003},
  doi     = {10.1016/S1071-5819(03)00038-7}
}

@article{lee2004trust,
  author  = {John D. Lee and Katrina A. See},
  title   = {Trust in Automation: Designing for Appropriate Reliance},
  journal = {Human Factors},
  volume  = {46},
  number  = {1},
  pages   = {50--80},
  year    = {2004},
  doi     = {10.1518/hfes.46.1.50_30392}
}

@article{parasuraman2010complacency,
  author  = {Raja Parasuraman and Dietrich H. Manzey},
  title   = {Complacency and Bias in Human Use of Automation: An Attentional Integration},
  journal = {Human Factors},
  volume  = {52},
  number  = {3},
  pages   = {381--410},
  year    = {2010},
  doi     = {10.1177/0018720810376055}
}

@inproceedings{tankelevitch2024metacognitive,
  author    = {Lev Tankelevitch and Viktor Kewenig and Auste Simkute and
               Ava Elizabeth Scott and Advait Sarkar and Abigail Sellen and
               Sean Rintel},
  title     = {The Metacognitive Demands and Opportunities of Generative AI},
  booktitle = {Proceedings of the 2024 CHI Conference on Human Factors in Computing Systems},
  year      = {2024},
  articleno = {680},
  pages     = {1--24},
  publisher = {Association for Computing Machinery},
  doi       = {10.1145/3613904.3642902}
}

@article{zhu2026meaningful,
  author  = {Liming Zhu and Qinghua Lu and Ming Ding and Sung Une Lee and Chen Wang},
  title   = {Designing Meaningful Human Oversight in AI},
  journal = {AI and Ethics},
  volume  = {6},
  articleno = {286},
  year    = {2026},
  publisher = {Springer},
  doi     = {10.1007/s43681-026-01147-7}
}

@article{deJong2025cognitive,
  author  = {Sander de Jong and Ville Paananen and Benjamin Tag and Niels van Berkel},
  title   = {Cognitive Forcing for Better Decision-Making: Reducing Overreliance on AI Systems Through Partial Explanations},
  journal = {Proceedings of the ACM on Human-Computer Interaction},
  volume  = {9},
  number  = {CSCW2},
  pages   = {1--30},
  year    = {2025},
  doi     = {10.1145/3710946}
}

@inproceedings{takerngsaksiri2025human,
  author={Takerngsaksiri, Wannita and Pasuksmit, Jirat and Thongtanunam, Patanamon and Tantithamthavorn, Chakkrit and Zhang, Ruixiong and Jiang, Fan and Li, Jing and Cook, Evan and Chen, Kun and Wu, Ming},
  title={Human-In-the-Loop Software Development Agents},
  booktitle={Proceedings of the 2025 IEEE/ACM 47th International Conference on Software Engineering: Software Engineering in Practice (ICSE-SEIP)},
  pages={342--352},
  year={2025},
  publisher={IEEE},
  doi={10.1109/ICSE-SEIP66354.2025.00036}
}

@inproceedings{chung2026collaborator,
  author={Chung, Young Jo and Hassan, Safwat},
  title={Collaborator or Assistant? How AI Coding Agents Partition Work across Pull Request Lifecycles},
  booktitle={Proceedings of the 3rd ACM International Conference on AI-Powered Software},
  pages={61--70},
  year={2026},
  publisher={Association for Computing Machinery},
  doi={10.1145/3805760.3814893}
}

@article{wang2026humans,
  author={Wang, Zora Z. and Yang, John and Lieret, Kilian and Tartaglini, Alexa and Chen, Valerie and Wei, Yuxiang and Wang, Zijian and Zhang, Lingming and Narasimhan, Karthik and Schmidt, Ludwig and Neubig, Graham and Fried, Daniel and Yang, Diyi},
  title={Humans are Missing from AI Coding Agent Research},
  journal={arXiv preprint arXiv:2608.12355},
  year={2026}
}

@inproceedings{dhanorkar2026oversight,
  author    = {Shipi Dhanorkar and Samir Passi and Mihaela Vorvoreanu},
  title     = {Human Oversight of Agentic Systems in Practice:
               Examining the Oversight Work, Challenges, and Heuristics
               of Developers Using Software Agents},
  booktitle = {Proceedings of the 2026 ACM Conference on Fairness,
               Accountability, and Transparency},
  pages     = {6438--6465},
  year      = {2026},
  publisher = {ACM},
  doi       = {10.1145/3805689.3812402}
}

@article{obrien2025more,
  title={More Code, Less Validation: The Impact of AI on Scientific Programming},
  author={O'Brien, Daniel and others},
  journal={arXiv preprint arXiv:2512.19644},
  year={2025}
}

@article{kaufman2026programmers,
  title={Programmers Are Poor and Overconfident Judges of LLM-Generated Assertions},
  author={Kaufman, Samuel and others},
  journal={arXiv preprint arXiv:2607.08885},
  year={2026}
}

@article{li2021measuring,
  author  = {Li, Shan},
  title   = {Measuring Cognitive Engagement: An Overview of Measurement Instruments and Techniques},
  journal = {International Journal of Psychology and Educational Studies},
  volume  = {8},
  number  = {3},
  pages   = {63--76},
  year    = {2021},
  doi     = {10.52380/ijpes.2021.8.3.239}
}

@article{dmello2010multimodal,
  author  = {D'Mello, Sidney K. and Graesser, Arthur},
  title   = {Multimodal Semi-Automated Affect Detection from Conversational Cues, Gross Body Language, and Facial Features},
  journal = {User Modeling and User-Adapted Interaction},
  volume  = {20},
  number  = {2},
  pages   = {147--187},
  year    = {2010},
  doi     = {10.1007/s11257-010-9074-4}
}

@article{kosch2023workload,
  author  = {Kosch, Thomas and Karolus, Jakob and Zagermann, Johannes and Reiterer, Harald and Schmidt, Albrecht and Wo{\'z}niak, Pawe{\l} W.},
  title   = {A Survey on Measuring Cognitive Workload in Human-Computer Interaction},
  journal = {ACM Computing Surveys},
  volume  = {55},
  number  = {13s},
  articleno = {283},
  pages   = {1--39},
  year    = {2023},
  publisher = {Association for Computing Machinery},
  doi     = {10.1145/3582272}
}

@article{dmello2009body,
  author  = {D'Mello, Sidney K. and Graesser, Arthur C.},
  title   = {Automatic Detection of Learner's Affect from Gross Body Language},
  journal = {Applied Artificial Intelligence},
  volume  = {23},
  number  = {2},
  pages   = {123--150},
  year    = {2009},
  doi     = {10.1080/08839510802631745}
}

@article{tao2019physiological,
  author  = {Tao, Da and Tan, Haibo and Wang, Hailiang and Zhang, Xu and Qu, Xingda and Zhang, Tingru},
  title   = {A Systematic Review of Physiological Measures of Mental Workload},
  journal = {International Journal of Environmental Research and Public Health},
  volume  = {16},
  number  = {15},
  pages   = {2716},
  year    = {2019},
  doi     = {10.3390/ijerph16152716}
}

@article{charles2019mental,
  author  = {Charles, Rebecca L. and Nixon, Jim},
  title   = {Measuring Mental Workload Using Physiological Measures: A Systematic Review},
  journal = {Applied Ergonomics},
  volume  = {74},
  pages   = {221--232},
  year    = {2019},
  doi     = {10.1016/j.apergo.2018.08.028}
}

@article{greene2015selfreport,
  author  = {Greene, Barbara A.},
  title   = {Measuring Cognitive Engagement With Self-Report Scales: Reflections From Over 20 Years of Research},
  journal = {Educational Psychologist},
  volume  = {50},
  number  = {1},
  pages   = {14--30},
  year    = {2015},
  doi     = {10.1080/00461520.2014.989230}
}

@inproceedings{stiber2024uh,
  title={" Uh, This One?": Leveraging Behavioral Signals for Detecting Confusion during Physical Tasks},
  author={Stiber, Maia and Bohus, Dan and Andrist, Sean},
  booktitle={Proceedings of the 26th International Conference on Multimodal Interaction},
  pages={194--203},
  year={2024}
}

% %%
% %% If your work has an appendix, this is the place to put it.
\appendix
\newpage
%TC:ignore
\section{Formative Interview Protocol}
\label{sec:app:formativeprotocol}

The formative study consisted of a 30-minute semi-structured interview conducted in person (10/12) or over Zoom (2/12). The interview explored participants' everyday use of AI assistants, moments of reduced oversight or evaluation, perceived consequences of such interactions, observable behavioral signals, and reactions to possible forms of re-engagement support. The questions below served as a guide; the interviewer used follow-up probes where appropriate.

\paragraph{1. Introduction, Consent, and Demographics}

The researcher introduced the study as an interview about people's experiences using AI assistants during everyday work. Participants were informed that there were no right or wrong answers and that the study focused on their own experiences and perceptions.

\begin{enumerate}
\item What AI tools do you currently use, and for what kinds of tasks?
\item How frequently do you use AI assistants?
\item What kinds of activities do you most often use them for (e.g., writing, coding, learning, brainstorming, information seeking, or decision-making)?
\end{enumerate}

\paragraph{2. Current AI-Assisted Interaction Practices}

Participants were asked to describe how they typically interact with AI assistants and how they evaluate or use AI-generated outputs.

\begin{enumerate}
\item Can you walk me through a recent example where you used an AI assistant for writing, coding, learning, brainstorming, information seeking, or decision-making?
\item When an AI gives you a response, what do you usually do with it?
\item How do you normally decide whether an AI-generated response is correct, useful, or appropriate?
\item Are there times when you inspect or evaluate the response carefully, and other times when you move forward with less checking? What tends to differ between those situations?
\end{enumerate}

\paragraph{3. Experiences of Reduced Oversight}

Participants were introduced to the idea that, during some AI-assisted interactions, users may continue progressing through a task while devoting relatively little attention to evaluating the AI's reasoning or output.

The interviewer provided the following description:

\begin{quote}
Sometimes people continue working with an AI assistant while doing relatively little checking or evaluation of what the AI has produced. They may accept, use, or move forward with an output without fully considering whether it is correct or appropriate for the task.
\end{quote}

Participants were then asked:

\begin{enumerate}
\item Does this description resonate with your own experiences using AI?
\item Can you recall a situation where you accepted or continued with an AI-generated output without evaluating it very closely?
\item What did that interaction look like?
\item Are there particular types of tasks where this is more likely to happen? 
\item What makes you more likely to reduce how much you check an AI response?
\item Do you always see this reduced level of oversight as negative, or can it sometimes be useful or appropriate?
\end{enumerate}

\paragraph{4. Perceived Consequences and Boundaries}

Participants discussed whether reduced oversight matters and how they distinguish problematic reliance from appropriate delegation.

\begin{enumerate}
\item Do you see reduced checking or evaluation of AI output as a problem? Why or why not?
\item What, if anything, can be lost when someone accepts AI output without evaluating it carefully?
\item Could this affect things such as learning, understanding, judgment, confidence, ownership, or control over the final outcome?
\item Are there situations where accepting an AI response with little evaluation is completely reasonable or useful?
\item How would you distinguish intentional delegation to an AI system from unintentionally failing to evaluate its output?
\item Does the importance of oversight depend on the task or the consequences of an error? How?
\end{enumerate}

\paragraph{5. Behavioral Signals and Re-engagement Support}

Participants were asked to reflect on what active versus reduced oversight might look like during interaction and how easily they could re-engage.

\begin{enumerate}
\item When you enter a period where you are checking the AI less carefully, how long does that usually last?
\item What usually causes you to start paying closer attention again?
\item How much effort does it take for you to re-engage with the task?
\item If someone were observing your interaction, what signs might suggest that you are actively evaluating the AI output?
\item What signs might suggest that you are accepting or moving forward with relatively little evaluation?
\item At what point during an interaction would a reminder to check the AI's output be most useful?
\item When would such a reminder feel unnecessary, annoying, or too late?
\end{enumerate}

% Participants were then presented with the following concept:

% \begin{quote}
% Imagine you are working with an AI assistant on a writing, learning, coding, or decision-making task. A wearable device, such as smart glasses, observes lightweight cues about your attention and interaction---for example, whether you inspected the AI response, compared it against other information, or quickly moved on. If the system detects that you may be accepting an output with relatively little evaluation, it provides a brief nudge intended to encourage you to check your understanding before continuing.
% \end{quote}

% Participants were asked:

% \begin{enumerate}
% \item What is your reaction to this idea?
% \item How would you feel about receiving this kind of nudge?
% \item Would a lightweight reminder be enough to make you re-engage, or would you prefer stronger support?
% \item For example, would asking you to explain your reasoning, inspect a particular part of the output, or verify an assumption be helpful?
% \item What would make such a system feel supportive rather than judgmental, distracting, or intrusive?
% \item Would you want the system to behave differently depending on the task, stakes, or reason you were relying on the AI?
% \end{enumerate}

\paragraph{6. Closing}

\begin{enumerate}
\item Is there anything else about how you evaluate or rely on AI-generated outputs that we have not discussed?
\item Do you have any concerns or suggestions regarding systems intended to support greater oversight during AI-assisted work?
\end{enumerate}

Participants were thanked for their time and participation.

\section{Programming Tasks}
\label{sec:app:protasks}

Table~\ref{tab:task-defects} explains the 6 different tasks and the two sets we used with their silent defect conditions.

\begin{table*}[h]
\centering
\caption{Programming tasks and defects used in the study, organized into two task sets. Silent defects produced incorrect implementations that passed the hidden pre-merge checks, whereas blatant defects caused the checks to fail.}
\label{tab:task-defects}
\small
\setlength{\tabcolsep}{4pt}
\renewcommand{\arraystretch}{1.15}
\begin{tabularx}{\textwidth}{
    >{\raggedright\arraybackslash}p{1.6cm}
    >{\raggedright\arraybackslash}X
    >{\raggedright\arraybackslash}X
    >{\raggedright\arraybackslash}X
    >{\raggedright\arraybackslash}X}
\toprule
\textbf{Task} &
\textbf{Function} &
\textbf{Silent defect} &
\textbf{What reveals the defect} &
\textbf{Blatant defect} \\
\midrule

\multicolumn{5}{l}{\textbf{Set 1}}\\
\addlinespace[2pt]

\textsc{Conflicts} &
Merges overlapping class time slots by room. &
Uses \texttt{next\_start <= current\_end} instead of \texttt{<}, causing back-to-back classes to be merged. &
Two classes sharing an exact boundary must remain separate. &
Returns \texttt{None} instead of the result list. \\

\textsc{Billing} &
Reconciles approved charges, refunds, and cancellations into member totals. &
Treats a refund as a cancellation, ignoring it instead of subtracting it from the total. &
An approved charge and an equal refund for the same member must produce a net total of zero. &
Returns an empty result despite receiving valid charges. \\

\textsc{Equipment} &
Determines each machine's current status from a log feed. &
Keeps the first reading for a repeated machine ID instead of the reading with the latest timestamp. &
An out-of-order duplicate must be resolved using its timestamp rather than its input position. &
Stores \texttt{None} as the status for every valid reading. \\

\midrule
\multicolumn{5}{l}{\textbf{Set 2}}\\
\addlinespace[2pt]

\textsc{Tags} &
Merges duplicate class tags case-insensitively and sorts the result. &
Initially merges tags correctly but reverts to case-sensitive matching during the final sorting step. &
Mixed-case duplicates, such as \texttt{HIIT} and \texttt{hiit}, must remain merged after sorting. &
Returns the input unchanged, leaving duplicates visibly separate. \\

\textsc{Attendance} &
Builds per-class rosters from check-in scans while removing duplicate check-ins. &
Uses a global set of seen members rather than tracking each class--member pair, dropping members who attend multiple classes. &
A member checked into two classes must appear once in each class roster. &
Returns an empty list despite receiving valid check-ins. \\

\textsc{Waitlist} &
Builds a promotion queue from waitlist requests while deduplicating members. &
Keeps the first encountered request for a repeated member instead of selecting the request with the earliest \texttt{joined\_at} value. &
An earlier request appearing later in the input must replace the first one encountered. &
Returns \texttt{None} instead of the promotion queue. \\

\bottomrule
\end{tabularx}
\end{table*}

\section{Performance per task}

Table~\ref{tab:task-difficulty-spread} explains how many people resolved each of the 6 different tasks under the silent condition.

\begin{table*}[h]
\centering
\caption{Silent-defect task outcomes for the 36-participant core cohort. Catch rate varies nearly tenfold across tasks (8\%--75\%), while self-reported confidence stays nearly flat (3.83--4.50 out of 5), motivating the task-controlled analyses used throughout the paper.}
\label{tab:task-difficulty-spread}
\begin{tabular}{lccccc}
\toprule
Task & $n$ & Caught & Catch rate & Mean confidence & Mean duration (min) \\
\midrule
CONFLICTS  & 12 & 1 & 8\%  & 4.08 & 8.67 \\
ATTENDANCE & 12 & 3 & 25\% & 4.08 & 7.30 \\
EQUIPMENT  & 12 & 4 & 33\% & 4.00 & 7.92 \\
TAGS       & 12 & 6 & 50\% & 3.83 & 10.11 \\
WAITLIST   & 11 & 5 & 45\% & 4.27 & 6.40 \\
BILLING    & 12 & 9 & 75\% & 4.50 & 6.14 \\
\midrule
Overall & 71 & 28 & 39\% & 4.13 & -- \\
\bottomrule
\end{tabular}
\end{table*}

%TC:endignore









\end{document}

\maketitle

\section{Supplementary Overview}

This document provides supplementary material for the paper, including additional methodological details, tables, figures, and supporting analyses.

\section{Feature Extraction}

This section details the feature extraction process. Section~\ref{app:telemetry-features} describes the feature extraction and processing for telemetry features, primarily keyboard and cursor recording data, while Section~\ref{app:physio-features} describes the feature extraction and processing for the physiological data.

\subsection{Telemetry Feature Extraction}
\label{app:telemetry-features}

This appendix documents the derivation of the telemetry input-log measures reported in the main paper, including the operational choices, thresholds, and quality gates
that the main text states only in summary. The measures themselves are listed in
Table~\ref{tab:features-telemetry}.

\subsubsection{Source and Constraints}
The interface logged keyboard and cursor activity as two event streams. Keystrokes were recorded as key-press events only, without corresponding key-release events, so hold time is not recoverable, and keystroke timing is characterized through press-to-press latency, combining hold and flight intervals. Cursor position was sampled at approximately 1\,Hz, so the straight-line displacement between consecutive samples is a lower bound on the path actually travelled, and sub-second corrective movements are not recoverable. This meant that we could not compute acceleration, jerk, angular velocity, curvature, direction-change rate, or path straightness, all of which require substantially higher sampling rates. Due to the limitations of the recording software, no click or scroll events were recorded, making click-based measures unavailable as well.

\subsubsection{Task Alignment}
The unit of analysis is a single main task, bounded by the timestamps at which the task brief was opened and the feature was handed off, excluding practice problems. The input logs recorded naive local time while session boundaries were held in UTC, and since participants were run across two timezones, we carried out alignment in UTC after verifying the clock offset against session boundaries for every participant. We also confirmed that both streams covered the full span of every analysed task, so no task is represented by a partial recording.

\subsubsection{Outcome Coding}
Analyses of oversight effectiveness use a per-task binary indicator of whether the silent defect was resolved at handoff. Each round yields a score out of three across its no-defect, silent-defect, and blatant-defect tasks. The no-defect implementation is correct as supplied, and the blatant-defect implementation must pass the visible pre-merge checks before a feature can be handed off, which leaves the silent-defect task as effectively the only one that can be handed off incorrect. This is why we treat a round score of three as indicating that the silent defect was resolved. Note that this is a measure of final correctness rather than of recognition, and the two can come apart. We use the term \emph{resolved} throughout rather than \emph{caught} for this reason.

\subsubsection{Operational Definitions}
Let $n$ be the number of key-press events within a task of duration $W$ seconds, and let $L_i = t_{i+1} - t_i$ be the press-to-press latencies. We define a \emph{pause} as a latency exceeding two seconds, and a \emph{burst} as a maximal run of keystrokes containing no latency exceeding one second. For the cursor, let $d_j$ be the Euclidean displacement between consecutive samples. A sample is \emph{idle} when $d_j = 0$, and an \emph{idle bout} is a maximal run of consecutive idle samples, whose length in samples approximates its length in seconds at the sampling rate used. We grouped keys into four categories: navigation keys (the four arrows, \texttt{HOME}, \texttt{END}, \texttt{PAGE UP}, \texttt{PAGE DOWN}), deletion keys (\texttt{BACKSPACE}, \texttt{DELETE}), shortcut modifiers (\texttt{META}, \texttt{CTRL}, \texttt{ALT}, left or right), and character keys, being everything that produces a character. We excluded shift and caps lock from the character category since they mostly reflect capitalisation rather than command use.

\subsubsection{Redundancy Among Measures}
Several of the 21 measures are strongly related by construction. The idle fraction and the moving-sample rate are near-complements ($\rho = -.99$), spatial spread and the horizontal standard deviation are near-duplicates ($\rho = .95$), and the two idle-bout measures correlate at $\rho = .91$. Given this redundancy, we still applied corrections for multiple comparisons over the 21 measures as listed rather than over a reduced set, which we considered the more conservative choice.

\subsubsection{Coverage and Missingness}
Timing, pause, and burst measures require at least five keystrokes within a task, and the within-burst rhythm measures require at least four latencies below the pause threshold. One of the 204 tasks fell below these thresholds, so it is missing those nine measures, but all other measures are complete for all 204 tasks. Task durations ranged from 0.57 to 33.48 minutes, with a median of 6.13.

\subsubsection{Threshold Choices}
Four thresholds are our own rather than inherited from prior work. These are the two-second pause boundary, the one-second burst boundary, the five-keystroke eligibility minimum, and the two-second cut-off separating within-burst rhythm from deliberation. Of these, the pause boundary is the consequential one, since the share of time inside pauses is the measure that carries the result reported in the main paper. Given how much rides on this choice, we repeated the outcome comparison across cut-offs of one, two, three, five, ten, and twenty seconds, and report the sensitivity in the main text.

\subsubsection{Analysis Procedures}
Pooled comparisons across tasks ignore the repeated-measures structure, so we report them only as descriptive. Within-participant comparisons use only participants who contributed both outcomes, and we tested these with a paired $t$-test, a Wilcoxon signed-rank test, and a sign-flip permutation test with 20{,}000 resamples. For the primary model, we decomposed each measure into a between-participant component, being that participant's mean, and a within-participant component, being the task deviation from that mean. We standardised both, and by construction the two are orthogonal. We fitted models as Bayesian mixed-effects logistic regressions with a participant random intercept, controlling for task duration, round, and position within the round. We report odds ratios with 95\% credible intervals, and $p$ values across families of measures are corrected with the Benjamini-Hochberg procedure and reported as $q$.

\subsubsection{Verification}
We recomputed keystroke rate and pause-time share for eight randomly selected tasks from the raw event files using an independent implementation, and found exact agreement. Event counts summed over tasks also reconcile with the raw files within the analysed spans. We checked that the within- and between-participant components were orthogonal and had the expected structure, and we checked the fixed-effect coefficient indexing in the mixed models against the design matrix column names, since a misalignment there would exchange the two components silently. We also checked the Benjamini-Hochberg implementation against a reference implementation. As an external check, we confirmed that participant-level task durations reproduce the pace annotations recorded during data collection.

\section{Gaze on the Study Interface}
\label{sec:supp:gaze}

The gaze measures reported in the main paper describe how participants' eyes moved, not what they moved over. Because gaze is expressed in the glasses' own frame, a fixation on the task brief and a fixation on the agent's response are indistinguishable. We therefore mapped each gaze sample onto a pixel of the participant's laptop display, using the first-person video to link the two streams, which let us attribute attention to a region of the interface and ask whether participants looked at the agent's output at all. We carried out this mapping on a sampled subset of sessions and report it here as a descriptive and methodological supplement, not as a study finding. The measures themselves are listed in Table~\ref{tab:features-gaze}.

\subsection{Source and Constraints}

The gaze stream and the screen recording come from separate devices, in different coordinate frames and on different clocks. Deriving screen-pixel attention from raw gaze therefore required the multi-stage pipeline described below, and the accuracy of the final mapping is bounded by its least accurate stage.

\subsection{Mapping Procedure}

The mapping has five steps. \emph{(1)~Time.} A full-screen sync marker was fired at the start and end of each session and appears in both the egocentric video and the screen recording. Matching the markers by pulse duration and spacing gives a linear map between the two clocks, accurate to $\pm0.10$~s; where both markers were recovered in both streams, the two offsets they imply agree to within 0.03~s over a session, and the fitted clock drift is indistinguishable from zero at this resolution. We confirmed each alignment visually on a frame in which the marker is simultaneously visible in the egocentric view and in the screen recording. \emph{(2)~Gaze ray.} We reconstructed the combined gaze direction in the glasses' central pupil frame and transformed it into the device frame using each participant's own factory calibration, which is device-specific and was never shared across participants. \emph{(3)~Screen pose.} Every 2~s we rectified the fisheye RGB frame to a pinhole image, matched it with SIFT against the screen frame displayed at that instant, and solved the display's 6-DoF pose with \texttt{solvePnP} over the inlier correspondences. Metric scale comes from the panel itself: the 14-inch display has a $302.4\times196.4$~mm active area, which the $1512\times982$ recording covers one-to-one. Because the laptop does not move, we express this pose in the odometry frame, where it is constant, and recompose a per-sample pose from the 800~Hz head trajectory. \emph{(4)~Intersection.} We intersected each gaze ray with the fitted display plane. We never reproject the gaze point at its reported vergence depth, since that depth disagrees with the range the plane implies by a median of $-8$ to $+27$~cm depending on the participant, with no consistent sign. We cannot correct for it, so we do not use it. \emph{(5)~Labels.} We detected the three interface panes from the median frame of each participant's own screen recording rather than assuming their layout. Across the sampled sessions the task brief is 279--282~px wide (56~mm), the editor 749--754~px (150~mm), and the agent pane 479--482~px (96~mm), with a header of 137--201~px. Task boundaries are brought in through a wall-clock bridge that ties the laptop's local clock to the recording to $\pm0.05$~s. The result is one row per 30~Hz gaze sample carrying a screen pixel, a pane, and a task.

\subsection{Verification}

On the sampled sessions the display pose was solved on 78--96\% of keyframes at a median reprojection error of 0.48--0.66~px. Two checks confirm the geometry without assuming a panel size: the aspect ratio recovered from the homography was 1.5394--1.5400 against the recording's 1.5397 ($\le0.02$\% error), and the scale factor that makes a static laptop sit still in the odometry frame was $k=0.985$--$1.018$. Splitting the keyframes in half and comparing the two independent fits puts the display-pose contribution at 0.74--1.47~px RMS (0.15--0.29~mm). The gaze model dominates this budget by roughly a factor of 40: the glasses' nominal $2^{\circ}$ accuracy propagates to 49--59~px, about 1--1.2~cm on the display, at median viewing ranges of 0.40--0.48~m. Time sync contributes nothing to the geometry, since the match is insensitive to a sync error of at least $\pm4$~s. A sync error instead corrupts which screen \emph{content} a sample is attributed to, not where on the panel it lands.

The panes are vertical columns 56, 150, and 96~mm wide, giving a 5--10$\times$ margin over the $1\sigma$ error: pane assignments survive a $1\sigma$ perturbation of the gaze ray 92.9--94.8\% of the time, and 67--75\% of samples sit more than $2\sigma$ from a column border. Attribution \emph{within} a pane, which line of code, which requirement, is not supported by this measurement, and we do not report it.

\subsection{Systematic Offset}

The mapping shows one systematic offset: the reconstruction sits low on the display, and 74--91\% of rays that hit the display plane but miss the panel miss below it. Estimating the offset against the mouse pointer, after dividing out the tendency of gaze and cursor to occupy the same pane, gives a downward offset of $+110$, $+50$, and $+10$~px in the three sessions carrying the full validation, and in two of the three the bootstrap interval includes zero. The sign is consistent, but the magnitude is not identified, and the estimator cannot separate a residual gaze-model bias from how far people look from their own pointer. We applied no correction, and instead recomputed each pane measure with the full estimated correction applied as a sensitivity check. Doing so moved no pooled pane share by more than 0.048. Because the offset is vertical and the panes are columns, it is largely orthogonal to the measure we use.

\subsection{Operational Definitions}

For each session and each task we computed the share of on-panel gaze time in each pane (task brief, editor, agent), the fraction of usable samples falling outside the display altogether, and two spatial descriptors: the standard deviation of the on-panel gaze position, and the entropy of its $20\times20$-bin histogram normalized to $[0,1]$, where a low value means gaze concentrated in one place and a high value means broad scanning. As elsewhere in our analysis, we compare conditions within participants rather than pooling across them.

\begin{figure*}[t]
  \centering
  \begin{minipage}{0.49\linewidth}
    \centering
    \includegraphics[width=\linewidth]{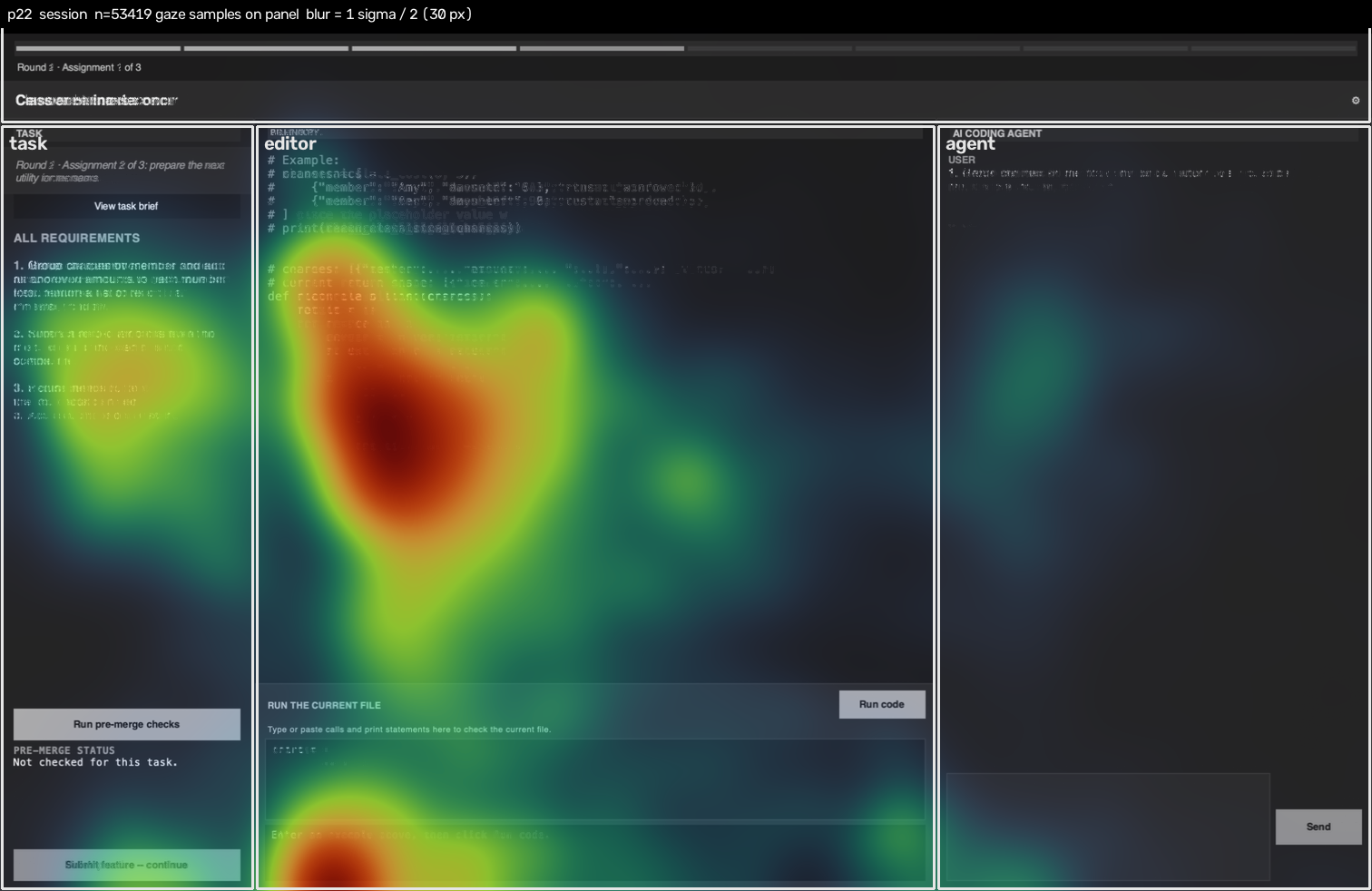}\\[2pt]
    {\small (a) P22: editor 0.73, agent 0.09, task brief 0.18}
  \end{minipage}\hfill
  \begin{minipage}{0.49\linewidth}
    \centering
    \includegraphics[width=\linewidth]{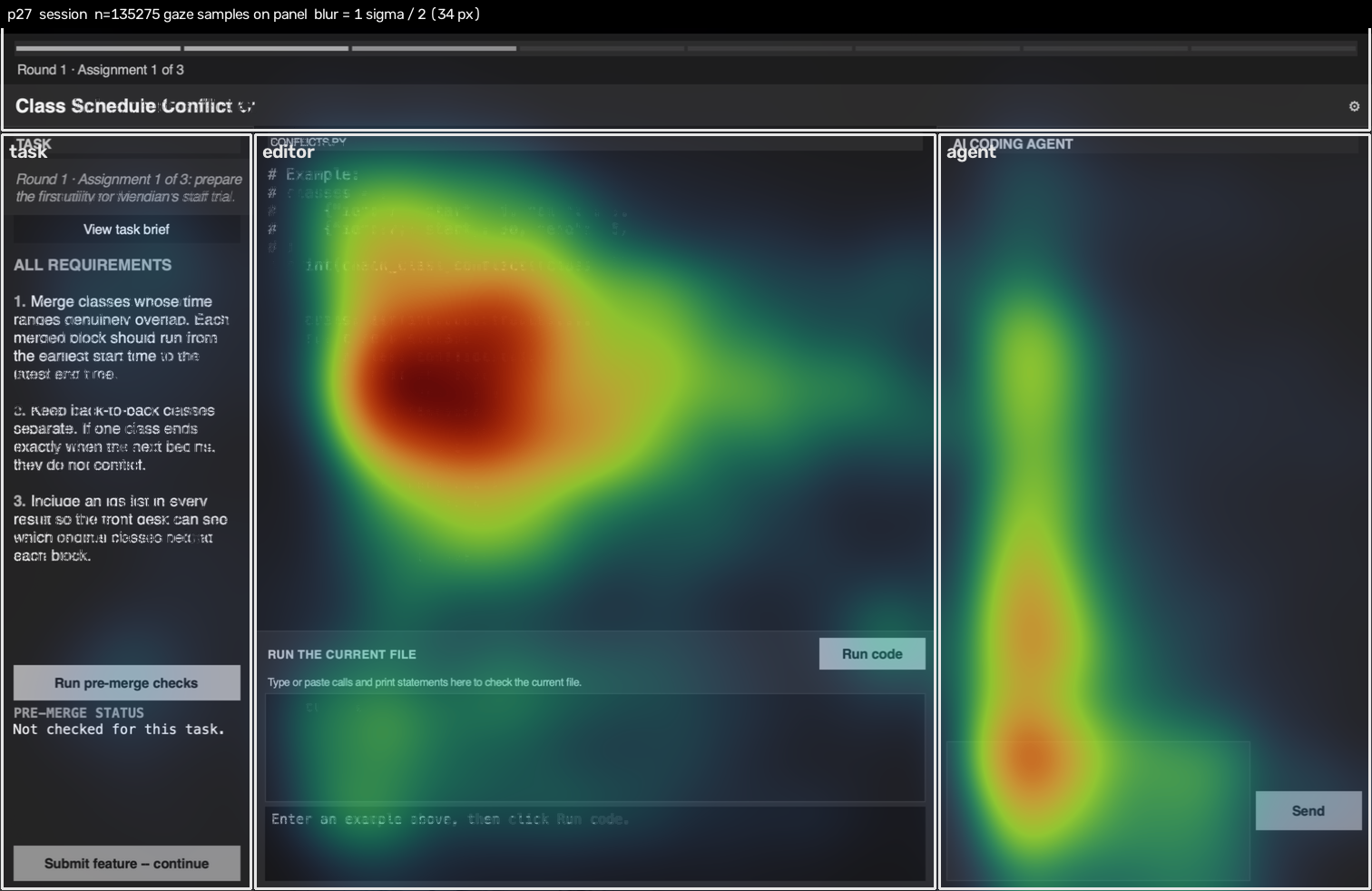}\\[2pt]
    {\small (b) P27: editor 0.65, agent 0.30, task brief 0.04}
  \end{minipage}\\[6pt]
  \begin{minipage}{0.49\linewidth}
    \centering
    \includegraphics[width=\linewidth]{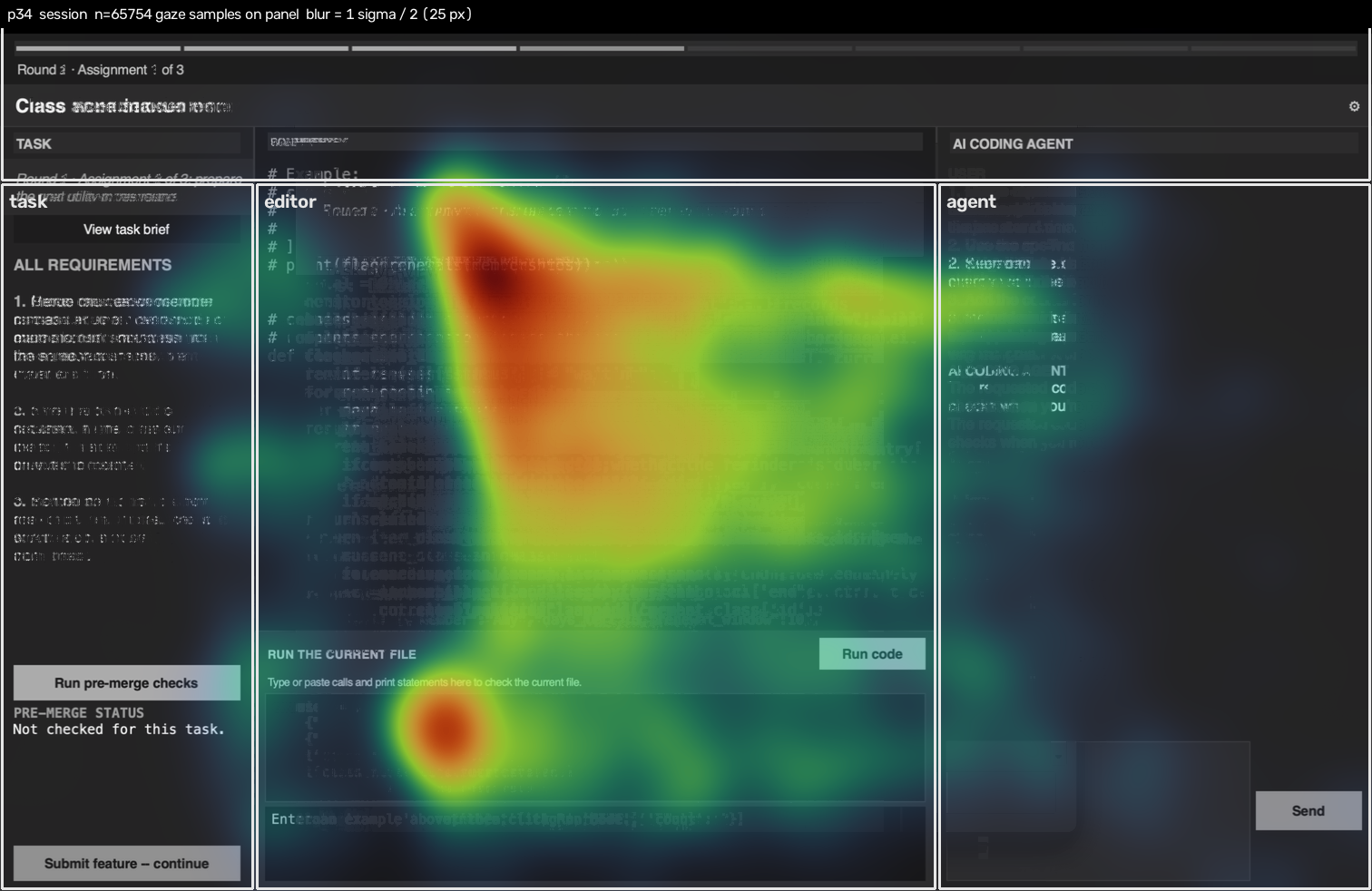}\\[2pt]
    {\small (c) P34: editor 0.83, agent 0.12, task brief 0.04}
  \end{minipage}\hfill
  \begin{minipage}{0.49\linewidth}
    \centering
    \includegraphics[width=\linewidth]{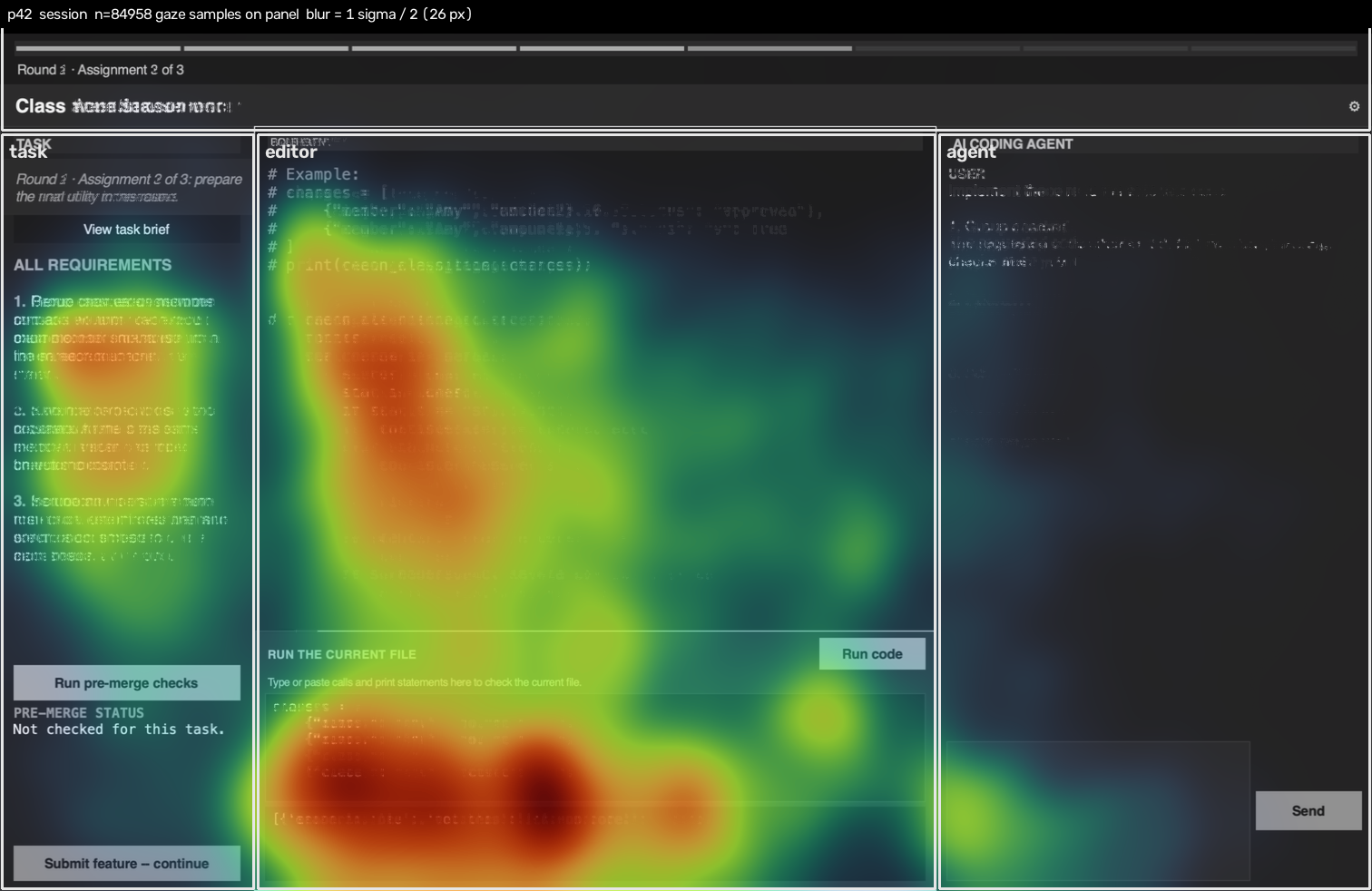}\\[2pt]
    {\small (d) P42: editor 0.76, agent 0.07, task brief 0.17}
  \end{minipage}
  \caption{Whole-session gaze heatmaps for four sampled participants, chosen to span the range of pane allocations we observed. Each map is drawn over the median frame of that participant's own screen recording, with the detected panes outlined; the values beneath each map are that session's pane shares. Density is smoothed with a Gaussian kernel set to half the session's $1\sigma$ mapping uncertainty (25--34~px) so that the picture cannot imply a precision the mapping does not have, and is normalized to each panel's own maximum, colors are therefore comparable within a panel but not between panels. The editor dominates every session, but the secondary structure differs: P27 reads down the agent column, P22 and P42 return repeatedly to the task brief and the manual-execution area, and P34 concentrates almost entirely on code and its own test inputs.}
  \label{fig:supp:heatmaps}
\end{figure*}

\subsection{Descriptive Results}

\textbf{Most gaze during a task was on the display, and most of it on the editor.} Across the sampled sessions we mapped 1.25~million gaze samples. A median of 86\% (range 68--98\%) of the usable samples fell on the display panel, and the remainder are rays that left the laptop entirely. Gaze was concentrated in the editor (median share 0.71, range 0.62--0.83), with less on the agent pane (0.17, range 0.07--0.30) and least on the task brief (0.06, range 0.03--0.18). Participants did direct some gaze at the agent's output, but the editor dominated the session for every participant we mapped. Figure~\ref{fig:supp:heatmaps} shows this for four of them.

\textbf{Defect condition was not reflected in where participants looked.} Comparing conditions within each participant across the sampled sessions with per-task boundaries ($n=10$ sessions, paired sign-flip permutation), agent-pane share on silent-defect tasks did not differ from no-defect tasks (mean difference $-0.003$, $p=.91$), and neither did editor-pane or task-pane share (both $p\ge.19$). Agent-pane share was nominally lower on blatant-defect than on no-defect tasks ($-0.058$, $p=.11$), which is the opposite of the direction that a warning-triggered increase in scrutiny would predict, and is not significant. This parallels the input-log and physiological results reported in the main paper: the conditions differed in how long a task took, but not in the gross allocation of gaze across the interface.

\textbf{Where participants looked reflected who they were more than what the task was.} The median within-participant range of agent-pane share across the three conditions was 0.052, against a between-participant spread of 0.066--0.296 in session agent-pane share ($SD=0.085$). The editor behaved the same way (0.065 within, 0.619--0.827 between). Absolute gaze measures aggregated at the grain of a whole task therefore function substantially as participant identifiers, consistent with the individual differences seen in the other multimodal measures. Agent-pane share also did not track silent-defect catch rate, either over the session ($\rho=.45$) or restricted to the silent-defect tasks themselves ($\rho=.15$). At this sample size this is not evidence of absence, but it gives no support for reading oversight quality off where a participant's gaze fell.

\subsection{Scope}

These results come from a sampled subset of sessions rather than the full cohort, each session's time alignment has to be verified by hand before its task labels can be used, and the comparisons above are within-participant with small numbers of sessions. We therefore treat pane-level attention as a measurement capability demonstrated here: interface-level attribution has a comfortable margin, but we make no finer-grained reading claims, and we do not treat these as condition- or outcome-level results.

\subsection{Physiological Feature Extraction}
\label{app:physio-features}

This appendix documents the derivation of the physiological measures reported in
the main paper. The measures themselves are listed in Table~\ref{tab:features-physio}. Task
alignment, outcome coding, and the statistical procedures follow
Appendix~\ref{app:telemetry-features}, with the one addition noted below.

\subsubsection{Source and Constraints}
Physiological measures derive from the wrist wearable worn for the duration of the session. Because the device exports processed streams rather than a raw photoplethysmography waveform, beat detection was performed on-device, and the timing of individual beats carries the manufacturer's filtering. This meant that pulse-wave morphology and pulse transit time were unavailable to us. Five streams were exported, being beat-to-beat intervals, heart rate at roughly 25-second cadence, a proprietary stress score, respiration rate, and per-minute actigraphy energy. We computed heart-rate variability from the beat-to-beat intervals and averaged the remaining streams within each task. All streams export UTC timestamps and were aligned to the task boundaries directly, without the offset correction required for the input logs.

\subsubsection{Artefact Handling}
We screened beat-to-beat intervals in two stages before computing any statistic. First, we rejected intervals outside a physiological plausibility range of 300 to 2000\,ms. We then passed the surviving intervals through a successive-difference criterion in the style of Malik, rejecting any interval departing by more than 20\% from the most recently accepted interval. We applied this criterion sequentially against the last accepted value rather than the last observed value, so that a single ectopic beat would not propagate through the series. The median rejection rate across usable tasks was 2.9\%, with a ninetieth percentile of 15.7\%.

\subsubsection{Measures Not Computed}
We did not compute frequency-domain measures, since the low-frequency band requires task lengths that a substantial proportion of tasks do not reach, and the resulting estimates would not have been comparable across tasks of unequal duration. Where the practice period yielded sufficient beats, we additionally computed a participant baseline over that period and expressed heart rate and the two log variability measures as changes from it. Baselines were available for 174 of 198 task records.

\subsubsection{Movement Covariate}
Because cardiac measures respond to physical activity, we carried per-minute actigraphy energy as a covariate in every physiological model, entered as $\log(1 + E)$, alongside task duration, round, and position within the round. We also separately confirmed that movement did not itself differ across defect conditions, which rules out the possibility that a cardiac effect was suppressed by an offsetting difference in movement.

\subsubsection{Quality Gates and Coverage}
We admitted a task into the analysis only when it contained at least 60 accepted beats and when accepted beats covered at least half the task span, where coverage is the summed duration of inter-beat gaps not exceeding five seconds, divided by task duration. Of the 34 participants in the input-log sample, one had no wearable recording at all. Of the remaining 198 task records, 180 met both gates, giving 32 participants with at least one usable task, 24 with all six, and 32 with at least four. Among the 18 excluded records, 14 failed the coverage gate and 6 the beat count. One participant contributed no usable task at all, since the recording contained a single gap spanning the entire session. Usable tasks contained between 65 and 1856 accepted beats, with a median of 330, and a median coverage of 0.90.

\subsubsection{Verification}
We recomputed mean interval length, mean heart rate, log RMSSD, and log SDNN for six randomly selected tasks from the raw interval files using an independent implementation of the same screening procedure, and found exact agreement. All extracted values fall within physiologically plausible ranges, heart rate spanning 53.5 to 111.0\,bpm across usable tasks, and the relation $SD1 \leq \mathrm{SDNN}$ holds throughout. As an end-to-end check on both timestamp alignment and the interval screening, we compared heart rate computed from the accepted intervals with the device's separately exported heart-rate stream over the same task windows. The two agree at $r = .993$ with a mean absolute discrepancy of 1.16\,bpm.

\subsubsection{Sample Differences Between Streams}
The two pipelines do not cover identical samples, so the counts reported in the main paper differ accordingly. The input-log analyses cover 34 participants and 204 tasks, while the physiological analyses cover 32 participants and 180 tasks. This difference arises from one participant without a wearable recording, one whose recording did not overlap the task period, and 18 individual tasks failing the coverage or beat count gates. Analyses that compare the two streams are restricted to tasks present in both.

% Telemetry feature table for Appendix \ref{app:telemetry-features}.
% Requires: booktabs.
\begin{table}[t]
\small
\label{tab:features-telemetry}
\begin{tabular}{@{}llp{0.34\linewidth}p{0.24\linewidth}@{}}
\toprule
\textbf{Group} & \textbf{Measure} & \textbf{Definition} & \textbf{Unit / note} \\
\midrule
\multicolumn{4}{@{}l}{\textit{Volume and composition}} \\
& Keystroke rate & $60n/W$ & keys/min \\
& Character ratio & Share of keys producing a character & proportion \\
& Deletion ratio & Share of \texttt{BACKSPACE}, \texttt{DELETE} & proportion \\
& Navigation ratio & Share of arrows, \texttt{HOME}, \texttt{END}, \texttt{PAGE UP/DOWN} & proportion \\
& Shortcut ratio & Share of \texttt{META}, \texttt{CTRL}, \texttt{ALT} & proportion \\
& Per-minute rates & Each category above, as a rate & keys/min; separates composition from volume \\
\addlinespace
\multicolumn{4}{@{}l}{\textit{Keystroke timing}} \\
& Median latency & $\mathrm{med}(L)$ & s \\
& Typing rhythm & $\mathrm{mean}(L_i \mid L_i < 2)$ & s; within-burst motor rhythm \\
& Rhythm CV & $\sigma / \mu$ of the same latencies & scale-free \\
\addlinespace
\multicolumn{4}{@{}l}{\textit{Pause and burst structure}} \\
& Pause rate & $60 \lvert \{i : L_i > 2\} \rvert / W$ & pauses/min \\
& Pause time share & $\sum_{L_i > 2} L_i / W$ & proportion of task \\
& Mean burst length & $n / (\lvert \{i : L_i > 1\} \rvert + 1)$ & keys/burst \\
& Longest gap & $\max(L)$ & s \\
\addlinespace
\multicolumn{4}{@{}l}{\textit{Cursor behaviour}} \\
& Idle fraction & Share of samples with $d_j = 0$ & proportion \\
& Travel distance & $60 \sum_j d_j / W$ & px/min; lower bound at 1\,Hz \\
& Movement rate & $60 \lvert \{j : d_j > 0\} \rvert / W$ & samples/min \\
& Mean idle bout & Mean run length of idle samples & s (approx.) \\
& Longest idle bout & Maximum run length of idle samples & s (approx.) \\
& Spatial spread & $\sqrt{\sigma_x^2 + \sigma_y^2}$ over positions & px \\
\bottomrule
\end{tabular}
\caption{Task-level measures extracted from the input logs. $n$ is the number
of key-press events in a task of duration $W$ seconds, $L_i$ the press-to-press latencies,
and $d_j$ the cursor displacement between consecutive samples. A pause is a latency above 2\,s, a burst a run of keystrokes containing no latency
above 1\,s, and an idle bout a run of consecutive cursor samples with zero displacement.}
\end{table}

% Gaze feature table for the gaze supplement (Section \ref{sec:supp:gaze}).
% Requires: booktabs.
\begin{table}[t]
\small
\label{tab:features-gaze}
\begin{tabular}{@{}llp{0.34\linewidth}p{0.24\linewidth}@{}}
\toprule
\textbf{Group} & \textbf{Measure} & \textbf{Definition} & \textbf{Unit / note} \\
\midrule
\multicolumn{4}{@{}l}{\textit{Gaze dynamics (glasses frame)}} \\
& Fixation rate & $60F/W$ & fixations/min \\
& Fixation duration & Mean duration of the $F$ fixations & ms \\
& Time fixating & Share of task time inside a fixation & proportion \\
& Saccade rate & $60S/W$ & saccades/min \\
& Saccade amplitude & Mean angular amplitude of the $S$ saccades & deg \\
& Gaze velocity & Mean angular change between consecutive samples, divided by the sample interval & deg/s \\
& Yaw dispersion & $\sigma$ of the combined gaze yaw & deg \\
& Pitch dispersion & $\sigma$ of the gaze pitch & deg \\
\addlinespace
\multicolumn{4}{@{}l}{\textit{Interface attention (mapped to the display)}} \\
& Pane share & Share of on-panel samples whose $(u,v)$ falls in the task-brief, editor or agent pane & proportion; three values per task \\
& Off-panel share & Share of usable gaze rays that do not land on the display panel & proportion; complement of the on-panel share \\
& Gaze dispersion & $\sigma_u$ and $\sigma_v$ of the on-panel gaze position & px \\
& Map entropy & Shannon entropy of the $20\times20$-bin histogram of on-panel $(u,v)$, divided by $\log_2 400$ & scale-free, $[0,1]$; low = concentrated, high = broad scanning \\
\bottomrule
\end{tabular}
\caption{Task-level measures extracted from the eye tracker. $F$ is the number of fixations
and $S$ the number of saccades in a task of duration $W$ seconds. The dynamics measures are
expressed in the glasses' own frame; the interface-attention measures are mapped to the
participant's laptop display, with $(u,v)$ denoting a gaze sample's on-panel screen position.}
\end{table}

% Physiological feature table for Appendix \ref{app:physio-features}.
% Requires: booktabs.
\begin{table}[t]
\small
\label{tab:features-physio}
\begin{tabular}{@{}llp{0.34\linewidth}p{0.24\linewidth}@{}}
\toprule
\textbf{Group} & \textbf{Measure} & \textbf{Definition} & \textbf{Unit / note} \\
\midrule
\multicolumn{4}{@{}l}{\textit{Cardiac}} \\
& Mean RR & $\overline{RR}$ & ms \\
& Mean HR & $60000 / \overline{RR}$ & bpm \\
& SDNN & $\sigma_{RR}$, log transformed & ms; right-skewed \\
& RMSSD & $\sqrt{\overline{\Delta RR^2}}$, log transformed & ms; right-skewed \\
& pNN50 & Share of $\lvert \Delta RR \rvert > 50$\,ms & proportion \\
& CVNN & $\sigma_{RR} / \overline{RR}$ & scale-free \\
& SD1 & $\sqrt{1/2}\,\sigma_{\Delta RR}$ & ms; Poincar\'e \\
& SD2 & $\sqrt{2\,\mathrm{SDNN}^2 - SD1^2}$ & ms; Poincar\'e \\
& SD2/SD1 & Ratio of the above & scale-free \\
\addlinespace
\multicolumn{4}{@{}l}{\textit{Other streams}} \\
& Stress score & Device value, averaged within task & 0--100; proprietary \\
& Respiration & Device value, averaged within task & breaths/min \\
& Movement energy & Actigraphy energy, averaged within task & entered as $\log(1+E)$ covariate \\
\bottomrule
\end{tabular}
\caption{Task-level measures extracted from the wearable. $RR$ denotes the accepted
beat-to-beat intervals within a task.}
\end{table}

% \section{Additional Methodological Details}

% Add supplementary methodology here.

% \section{Additional Tables}

% \begin{table}[ht]
%     \centering
%     \caption{Example supplementary table.}
%     \label{tab:supp-example}
%     \begin{tabular}{lll}
%         \toprule
%         Column 1 & Column 2 & Column 3 \\
%         \midrule
%         A & B & C \\
%         D & E & F \\
%         \bottomrule
%     \end{tabular}
% \end{table}

% \section{Additional Figures}

% \begin{figure}[ht]
%     \centering
%     % \includegraphics[width=\linewidth]{figures/example.pdf}
%     \caption{Example supplementary figure.}
%     \label{fig:supp-example}
% \end{figure}

% \section{Additional Results}

% Add supplementary analyses or results here.

% If you have a bibliography:
% \bibliographystyle{ACM-Reference-Format}
% \bibliography{references}